\documentclass[prd,preprintnumbers,aps,preprint,amsmath,amssymb,superscriptaddress,floatfix,nofootinbib,unsortedaddress]{revtex4}
\pdfoutput=1

\usepackage{amssymb, bm, amsmath, graphicx, physymb}
\usepackage[colorlinks=true,linktocpage=true,linkcolor=blue,citecolor=blue]{hyperref}
\usepackage[usenames,dvipsnames]{color}
\usepackage{slashed}
\usepackage[utf8]{inputenc}
\usepackage[T1]{fontenc}
\usepackage{enumerate}
\usepackage{amsfonts}
\usepackage{mathtools}
\usepackage{latexsym}
\usepackage{hyperref}
\usepackage{epstopdf}
\usepackage{bbm}
\usepackage{caption}
\usepackage{soul}
\usepackage[dvipsnames]{xcolor}
\usepackage[b]{esvect}

\newcommand{\tvec}[1]{\vec{#1}}

\renewcommand\a{\alpha}
\renewcommand\b{\beta}
\renewcommand\d{\delta}

\renewcommand\l{\lambda}
\renewcommand\r{\rho}

\newcommand\e{\epsilon}
\newcommand\g{\gamma}

\newcommand\m{\mu}

\newcommand\p{\pi}

\newcommand{\N}{\mathcal{N}}

\renewcommand{\vec}{\boldsymbol}
\renewcommand{\part}{{\rm part}}

\newcommand{\be}{\begin{equation}}
\newcommand{\ee}{\end{equation}}
\newcommand{\bes}{\begin{subequations}}
\newcommand{\ees}{\end{subequations}}
\newcommand{\bea}{\begin{eqnarray}}
\newcommand{\eea}{\end{eqnarray}}

\newcommand{\pa}{\partial}

\newcommand{\na}{\nabla}

\begin{document}

\title{Jet Broadening in Flowing Matter -- Resummation}

\author{Carlota Andres}
\email[Email: ]{carlota.andres-casas@polytechnique.edu}
\affiliation{CPHT, CNRS, \'Ecole polytechnique,
IP Paris, F-91128 Palaiseau, France}
\author{Fabio Dominguez}
\email[Email: ]{fabio.dominguez@usc.es}
\affiliation{Instituto Galego de F{\'{i}}sica de Altas Enerx{\'{i}}as (IGFAE),  Universidade de Santiago de Compostela, Santiago de Compostela 15782,  Spain}
\author{Andrey V. Sadofyev}
\email[Email: ]{andrey.sadofyev@usc.es}
\affiliation{Instituto Galego de F{\'{i}}sica de Altas Enerx{\'{i}}as (IGFAE),  Universidade de Santiago de Compostela, Santiago de Compostela 15782,  Spain}

\author{Carlos A. Salgado}
\email[Email: ]{carlos.salgado@usc.es}
\affiliation{Instituto Galego de F{\'{i}}sica de Altas Enerx{\'{i}}as (IGFAE),  Universidade de Santiago de Compostela, Santiago de Compostela 15782,  Spain}

\begin{abstract}
In this work, we obtain the leading subeikonal  corrections to the jet momentum broadening distribution in a QCD medium arising from the transverse flow of the matter. We first derive the single-particle propagator of a highly energetic parton resumming its multiple interactions with the homogeneous flowing matter, explicitly keeping the leading subeikonal flow terms. Then, we use this propagator to obtain the jet broadening distribution and its leading moments. We show that this distribution becomes anisotropic in the presence of transverse flow, since its odd moments are generally non-zero and proportional to the transverse velocity of the medium. Finally, we evaluate several odd moments, which we compare to the corresponding results at first order in opacity, showing that accounting for multiple in-medium scatterings is essential to describe some observables in dense nuclear matter.
\end{abstract}

\maketitle

\section{Introduction}

One of the most important signatures of the formation of the quark-gluon plasma (QGP) in high-energy nuclear collisions is the suppression of highly energetic particles, a phenomenon commonly known as jet quenching \cite{PHENIX:2001hpc,STAR:2002ggv}.  Although the suppression of such high-energy particles has been largely explained in terms of the energy loss due to their interactions with the QGP, recent developments both from theory and experiment have shown that the inner structure of the jets created by these particles is also significantly modified. Due to the sensitivity to all these effects, jets in heavy-ion collisions provide a unique tool to study the properties and structure of the QGP. This concept of jet tomography has attracted a great deal of attention in the literature, see for instance \cite{Vitev:2002pf, Majumder:2006wi, Xu:2014ica, Djordjevic:2016vfo, Apolinario:2017sob, Andres:2019eus, Feal:2019xfl, He:2020iow,  Apolinario:2020uvt, Sadofyev:2021ohn, Du:2021pqa, Antiporda:2021hpk,Hauksson:2021okc,Barata:2022krd,Fu:2022idl,Sadofyev:2022hhw} and references therein. 

The interaction of high-energy colored particles with the QGP is usually described within perturbative Quantum Chromodynamics (pQCD) as a scattering process where the medium is modelled with a background stochastic color field, see e.g.~\cite{Baier:1996kr, Baier:1996sk, Zakharov:1996fv, Zakharov:1997uu, Gyulassy:2000fs, Gyulassy:2000er, Wiedemann:2000za, Wang:2001ifa,Arnold:2002ja,Djordjevic:2003zk,Mehtar-Tani:2006vpj,Caron-Huot:2010qjx,Sievert:2018imd,Andres:2020vxs,Barata:2021wuf}. In this picture, highly energetic partons undergo multiple scatterings with the QCD matter, resulting in both the broadening of their transverse\footnote{Here and throughout this manuscript, ``transverse'' means orthogonal to the direction of the propagation of the high-energy parton.} momentum distribution and the emission of real radiation,  also referred to as medium-induced radiation. While this approach is rather general, several approximations are  made in order to simplify the calculations while capturing the main physical effects, e.g.~\cite{Casalderrey-Solana:2007knd,Sievert:2018imd,Blaizot:2012fh}. In this paper, we will focus on the high-energy approximation, known also as eikonal approximation, where all transverse momenta are considered smaller than the energy of the final parton $E$. This assumption implies that the broadening and radiation dynamics decouple from the transverse structure and evolution of the medium, as explained in \cite{Sadofyev:2021ohn}.

At this point, it is important to make a clarification about the use of this high-energy approximation in different settings and what is usually understood by the eikonal limit. On its most strict form, the eikonal limit refers to the case where all contributions of order $\mathcal{O}(\bot/E)$, where $\bot$ stands for any of the relevant transverse scales, are neglected. This is the eikonal limit used for instance in Color Glass Condensate (CGC) calculations, where the transverse positions of incoming partons are frozen during the interaction with the target and its wave function is only modified through a color rotation. For calculations involving in-medium emissions this approximation is too restrictive and non-trivial results are achieved only when terms scaling as $\frac{\bot^2}{E}z$ are kept to all orders, with $z$ a longitudinal position taking values up to the length of the medium, see e.g. \cite{Gyulassy:2000fs, Gyulassy:2000er,Wiedemann:2000ez}. These terms include phases arising from vacuum propagators in between scatterings, usually referred to as Landau-Pomeranchuk-Migdal (LPM) phases, which are essential to get the correct emission spectrum and yield to the well-known LPM suppression in the multiple scattering calculations. This is the case that will be used as the zeroth order for the calculations in this paper, and therefore we will refer to it as the eikonal limit, in contrast to the subeikonal contributions which scale as $\bot/E$ with no explicit length enhancement.

Since jets decouple from the transverse structure and dynamics of the QCD matter in the eikonal limit, these medium effects can only be included into jet-quenching formalisms by relaxing the eikonal approximation. Due to the difficulties inherent to such beyond-eikonal calculations, several works attempted to account for transverse flow effects focusing on the medium dilution effects \cite{Baier:1998yf, Gyulassy:2000gk, Gyulassy:2001kr}, using phenomenological motivated models \cite{Gyulassy:2001kr, Armesto:2004pt, Armesto:2004vz}, or considering purely kinematic arguments \cite{Baier:2006pt, Liu:2006he, Renk:2006sx}. It was not until very recently that matter transverse structure and flow were rigorously included in a pQCD description of in-medium radiation and broadening \cite{Sadofyev:2021ohn}, where the flow effects were shown to enter at subeikonal order. This calculation was performed at leading order in the Gyulassy-Levai-Vitev (GLV) opacity expansion framework \cite{Gyulassy:2000fs, Gyulassy:2000er,Wiedemann:2000za}, where only one scattering with the medium is allowed. Extending this result to account for all multiple in-medium scatterings by means of the Baier, Dokshitzer, Mueller, Peigné, Schiff, and Zakharov (BDMPS-Z) framework \cite{Baier:1996kr,Baier:1996sk,Zakharov:1996fv,Zakharov:1997uu} is then the next natural step in the development of this theory. This program was already started in \cite{Barata:2022krd} where the transverse gradients of matter properties were taken into account for the calculation of the broadening distribution.

In this work, we focus on the subeikonal corrections to the broadening distribution due to the transverse flow of the QGP. We first compute the in-medium propagator of a highly energetic parton resumming its multiple interactions with homogeneous flowing matter. Then, we use that result to derive the leading subeikonal corrections to the momentum broadening due to the transverse velocity of the medium, generalizing the result of \cite{Sadofyev:2021ohn} to all orders in opacity. We show that the all-order broadening distribution, isotropic in the static case, becomes anisotropic due to these transverse flow terms. While we focus here on the case of the QGP, similar effects could also arise in cold nuclear matter due to the collective motion of the nucleons.

The paper is organized as follows: in section~\ref{sec:propagator} we present the derivation of the amplitude of a high-energy parton moving throughout flowing homogeneous matter in the BDMPS-Z formalism. We obtain in section~\ref{sec:broadening} the broadening distribution by computing the average over the scattering centers of the squared amplitude. In section~\ref{sec:moments} we compute the odd moments of the final state momentum distribution, which in contrast with the static case do not vanish, thus probing that jet broadening is anisotropic in the presence of transverse flow. Finally, we summarize and conclude in section~\ref{sec:conclusions}.

\section{Resummation at the amplitude level: the propagator}
\label{sec:propagator}

In this section, we show how to obtain the in-medium amplitude of a fast moving parton with final energy $E$ in flowing homogeneous matter, accounting for multiple parton-medium interactions. We will express this amplitude in the form of a single-particle propagator, as usually done in the BDMPS-Z formalism, keeping the leading subeikonal corrections $\mathcal{O}(\bot/E)$ to capture the flow effects but disregarding effects of spin polarization by using scalar QCD for the interactions involving the leading parton, as done at first order in opacity in \cite{Sadofyev:2021ohn}.

\begin{figure}[t]
    \vspace{-1cm}
    \centerline{
    \includegraphics[scale=0.23]{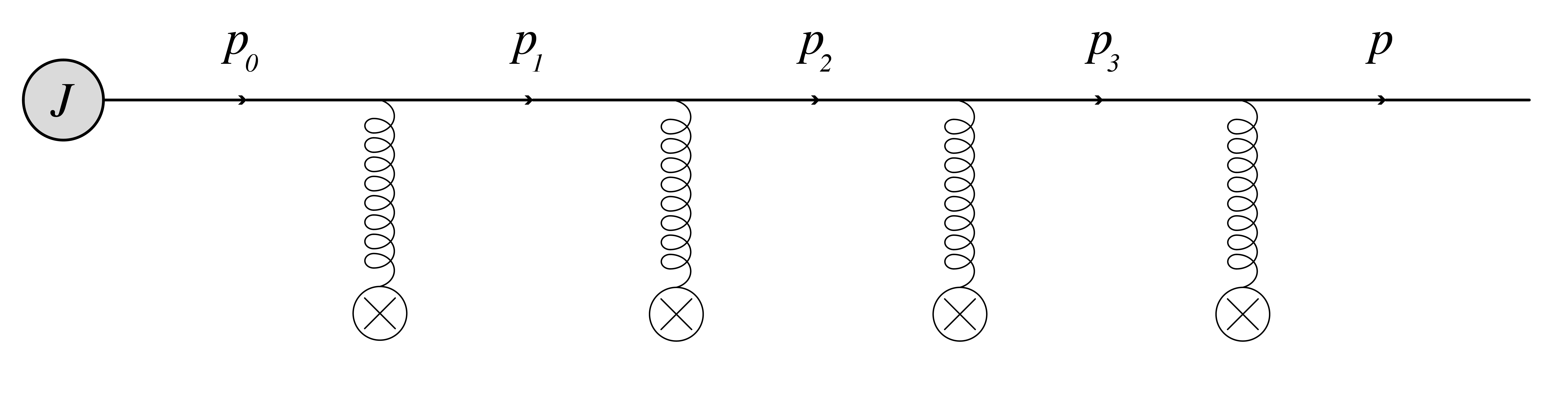}}
    \vspace{-0.5cm}
    \caption{The contribution $\mathcal{M}_4$ with four in-medium scatterings to the full amplitude, where the final state momentum is on-shell $p=\left(E, \vec{p}, E- \frac{p_\bot^2}{2E}\right)$,  see the details below.}
\label{fig:Mr}
\end{figure}

We consider the amplitude of the process in which the highly-energetic parton is created in the medium by some source $J(p)$, and then interacts $n$ times with the matter which is modeled with a background field, as shown in Fig.~\ref{fig:Mr}. The corresponding matrix element can be written as
\begin{align}
i\mathcal{M}_n(p)&=\int \prod_{j=0}^{n-1}\,\left[\frac{d^4p_j}{(2\p)^4}\,ig A_\mu(p_{j+1}-p_j)(p_{j+1}+p_j)^\mu\,\frac{i}{p_j^2+i\e}\,\right]\,J(p_0)\,,
\label{eq:Mn1}
\end{align}
where $p_n\equiv p$ is the final four-momentum of the parton, $A_\mu(q)=A_\mu^a(q)\,t^a$ is the background field, the index $\m$ runs over 4D space-time, $t^a$ is the color matrix in the representation of the projectile (the highly energetic parton), and the product is ordered from right to left with the $n$-th field insertion being the leftmost factor. 

Now, following \cite{Sadofyev:2021ohn}, the medium velocity can be included in the calculation by modifying the background field entering Eq.~(\ref{eq:Mn1}). In order to do so, we model the matter as a collection of massive color sources moving with non-relativistic velocity\footnote{Throughout this manuscript bold font will be used for 2D vectors in the transverse plane with respect to the leading parton large momentum component $p_z$.} $u^\mu=(1, \vec{u},u_z)$, as explained in detail in Section II C of \cite{Sadofyev:2021ohn}. We introduce here a subtle modification with respect to \cite{Sadofyev:2021ohn}: we make use of the color charge density $\hat\rho^a(\vec{x},z)$ in coordinate space, allowing us to have a continuous distribution of medium sources (which will be averaged over after squaring the amplitude), as is the common practice both in the CGC and BDMPS-Z formalisms. We also set the longitudinal velocity equal to zero $u_z=0$, since its effects can be obtained by performing a longitudinal boost. We focus, instead, on calculating the subeikonal corrections arising from the transverse velocity $\vec{u}$, which cannot be recovered by a transverse boost, since transverse boosts do not commute with the eikonal expansion. Hence, the background field used in our calculation has the following form
\be
\label{eq:A_mu}
g A^{a\l}(q) = u^\l\, v(q)\,\left[\int d^2\tvec{x}\,dz\,e^{-i\left(\tvec{q}\cdot\tvec{x}+q_z z\right)}\hat\rho^a(\tvec{x}, z)\right]\,(2\p)\,\d(q^0-\vec{u}\cdot \vec{q})\,,
\ee
where $v$ is the interaction potential, and we have neglected the recoil of the sources. We rely on the  
Gyulassy-Wang (GW) model \cite{Gyulassy:1993hr}, and thus set $v$ to
\be
v(q) = \frac{g^2}{q^2-\mu^2+i\epsilon}\,,
\label{eq:potential}
\ee
where $\m$ is the Debye mass of the QGP or another characteristic screening scale (e.g. in the case of cold nuclear matter).

Note that the field of a discrete collection of scattering centers used in \cite{Sadofyev:2021ohn}, can be straightforwardly obtained from Eq.~(\ref{eq:A_mu}) by taking $\hat\r^a(\vec{x}, z)=\sum_j t_j^a\, \d^{(2)}(\vec{x}-\vec{x}_j)\,\d(z-z_j)$, where $t_j^a$ is the color generator of $j$th source.

Plugging \eqref{eq:A_mu} into \eqref{eq:Mn1} we can easily perform the integrals over the zero components of all the momenta. Then, we perform all the integrals over the $z$-components using contour integration, where only one type of poles of the intermediate propagators of the fast moving parton contributes to the desired accuracy, as explained for instance in \cite{Sadofyev:2021ohn,Casalderrey-Solana:2007knd}. Indeed, since we are considering only highly energetic partons, a single interaction is not able to change the sign of the large component of the projectile momentum, thus fixing the sign of $p_{jz}$ for all $j$ up to higher subeikonal orders. Notice also that the potentials $v(q)$ are screened and the corresponding poles are suppressed under the assumption that the sources are sufficiently separated. Then, the zero and longitudinal components of the intermediate momenta are set to
\begin{subequations}
\label{p0z}
\begin{align}
p_{j0} &\simeq u\cdot p +\vec{u}\cdot\vec{p}_j\, ,\\
p_{jz} &\simeq p_{j0} - \frac{p_{j\bot}^2}{2E} = u\cdot p +\vec{u}\cdot\vec{p}_j- \frac{p_{j\bot}^2}{2E}\, ,
\end{align}
\end{subequations}
where $u\cdot p\equiv u^\m p_\m\simeq E-\vec{u}\cdot\vec{p}$, $p^0 \equiv E$ is the parton final energy, $p_{j\bot}\equiv|\vec{p}_j|$ is the magnitude of the 2D vector $\vec{p}_j$, and we have kept just the  zeroth and first order terms in the eikonal expansion. 

Now, upon reshuffling some of the Fourier factors, the amplitude reduces to
\begin{align}
i\mathcal{M}_n(p)= & \int \prod_{j=0}^{n-1}\,\Bigg[dz_{j+1}\,\frac{d^2\vec{p}_{j}}{(2\p)^2}\, i\left(1 - \frac{\vec{u} \cdot \vec{p}_{j}}{E}\right)\,v(\tilde{p}_{j+1}-\tilde{p}_j) \nonumber\\
&\times\hat\rho^{a_{j+1}}(\vec{p}_{j+1}-\vec{p}_{i},z_{j+1})t^{a_{j+1}}\theta(z_{j+1}-z_j) \,e^{i\left(\vec{u} \cdot \vec{p}_{j}-\frac{p_{j\bot}^2}{2E}\right)\left(z_{j+1}-z_j\right)} \Bigg]\nonumber \\
&\times e^{-i\left(\vec{u} \cdot \vec{p}-\frac{p_{\bot}^2}{2E}\right)z_n}\,J(\tilde{p}_0)\,,
\label{eq:Mn2d}
\end{align}
where we have defined $\tilde{p}_j$ as the four-momentum of the leading parton with its zero and longitudinal components replaced according to Eq.~\eqref{p0z},  $\tilde{p}_j = \left(u\cdot p + \vec{u}\cdot\vec{p}_{j},\vec{p}_{j},u\cdot p+\vec{u}\cdot\vec{p}_{j}-\frac{p_{j\bot}^2}{2E}\right)$, $t^{a_j}$ is the color generator of the leading parton after the $j$th interaction (not to be confused with $t^a_j$), and, without loss of generality, we have set $z_0 = 0$. Here we have also used the mixed representation of the color charge operator where only the transverse components are Fourier transformed while the longitudinal coordinate is explicitly kept,
\be
\hat\rho^a(\vec{q},z)
= \int d^2\vec{x}\, e^{-i\vec{q}\cdot\vec{x}}\,
\hat \rho^a(\vec{x},z)\,.
\ee
Notice that the LPM phases in (\ref{eq:Mn2d}) are affected by the flow velocity, as shown in \cite{Sadofyev:2021ohn}. 

At this point, we can write the amplitude as a convolution of the initial source $J$ with an in-medium propagator ${\cal G}(\vec{p},L; \vec{p}_{0},z_0)$ in the following way:
\bea
\label{eq:amplitudeprop}
i\mathcal{M}_n(p)
=\,
e^{-i\left(\vec{u}\cdot\vec{p}-\frac{p_{\bot}^2}{2E}\right)L}
\int 
\frac{d^2\vec{p}_{0}}{(2\p)^2} \, \mathcal{G}_n(\vec{p},L; \vec{p}_{0},0)
\,J(\tilde{p}_0)\,,
\eea
where the $n$th-order perturbative contribution to the propagator is given by
\begin{align}
\mathcal{G}_n(\vec{p},L; \vec{p}_{0},z_0)\, =&
\prod_{j=1}^n\,
\Bigg[
\int_{z_0}^L dz_j\,
\int\frac{d^2\vec{p}_{j}}{(2\p)^2}
\,i 
\left(1 - \frac{\vec{u} \cdot \vec{p}_{j-1}}{E}\right)
\,v(\tilde{p}_j-\tilde{p}_{j-1})\,
\hat\rho^{a_j}(\vec{p}_{j}-\vec{p}_{j-1},z_j)\,
t^{a_j} 
\nonumber\\
&\hspace{-1.5cm}\times
\theta(z_j-z_{j-1})
\,
e^{i\left(\vec{u} \cdot \vec{p}_{j-1}-\frac{p_{j-1\bot}^2}{2E}\right)\left(z_{j}-z_{j-1}\right)}
\Bigg]
\, (2\pi)^2
\delta^{(2)}(\vec{p}-\vec{p}_n)
\,e^{i\left(\vec{u} \cdot \vec{p}-\frac{p_{\bot}^2}{2E}\right)(L-z_n)}\,.
\label{eq:Gn}
\end{align}

Before turning to the resummation procedure, it is useful to cast \eqref{eq:Gn} as a recursion relation in the following way
\begin{align}
\mathcal{G}_n(\vec{p},L; \vec{p}_{0},z_0)\, = 
i
\int_{z_0}^L dz \int\frac{d^2\vec{l}}{(2\p)^2}
\,\left(1 - \frac{\vec{u} \cdot \vec{l}}{E}\right)
\,v(\tilde{p}-\tilde{l})
\,\hat\rho^a(\vec{p}-\vec{l},z)
\,t^a \notag
\\
\times
\, e^{i\left(\vec{u} \cdot 
\vec{p}-\frac{p_{\bot}^2}{2E}\right)(L-z)} 
\,\mathcal{G}_{n-1}(\vec{l},z; \vec{p}_{0},z_0)\,,
\label{eq:gn_recrusive}
\end{align}
where $\tilde l=\left(u\cdot p+\vec{u}\cdot \vec{l}, \vec{l}, u\cdot p+\vec{u}\cdot\vec{l}-\frac{l_\bot^2}{2E}\right)$. Now Eq.~(\ref{eq:gn_recrusive}) can be straightforwardly summed over $n$ from zero to infinity, leading to an integral equation for the full propagator
\begin{align}
\mathcal{G}(\vec{p},L; \vec{p}_{0},z_0)
= 
\mathcal{G}_0(\vec{p},L; \vec{p}_{0},z_0) 
+i
&\int_{z_0}^L dz
\int\frac{d^2\vec{l}}{(2\p)^2} 
\,\left(1 - \frac{\vec{u} \cdot \vec{l}}{E}\right)
\,v(\tilde{p}-\tilde{l})
\,\hat\rho^a(\vec{p}-\vec{l},z)
\,t^a 
\notag 
\\
\times 
\, &
e^{i\left(\vec{u} \cdot \vec{p}-\frac{p_{\bot}^2}{2E}\right)(L-z)} \,\mathcal{G}(\vec{l},z; \vec{p}_{0},z_0)\,,
\label{eq:propSchwDy}
\end{align}
where, by focusing on the case of no interactions, one can see that the correct initial propagator  $\mathcal{G}_0(\vec{p},L; \vec{p}_{0},z_0)$ is given by
\be
\mathcal{G}_0(\vec{p},L; \vec{p}_{0},z_0)
= 
(2\pi)^2\delta^{(2)}
(\vec{p}-\vec{p}_{0})\,
e^{i\left(\vec{u} \cdot \vec{p}-\frac{p_{\bot}^2}{2E}\right)(L-z_0)}\,.
\ee

The integral equation above can also be 
re-written as a differential equation in $L$, which reads
\begin{align}
\frac{\partial}{\partial L}\mathcal{G}(\vec{p},L; \vec{p}_{0},z_0) 
&= 
i\left(\vec{u} \cdot \vec{p}-
\frac{p_{\bot}^2}{2E}\right)
\mathcal{G}(\vec{p},L; \vec{p}_{0},z_0)
\nonumber
\\
&\quad+i \,
\int\frac{d^2\vec{l}}{(2\p)^2} 
\,\left(1 - \frac{\vec{u} \cdot \vec{l}}{E}\right)
\,v(\tilde{p}-\tilde{l})
\,\hat\rho^a(\vec{p}-\vec{l},L)
\,t^a 
\,\mathcal{G}(\vec{l},L; \vec{p}_{0},z_0)\,.
\label{eq:diffeqprop}
\end{align}
It is clear that in the limit of zero transverse flow velocity this equation reduces to its standard form, see for instance \cite{Blaizot:2015lma}.

Besides the explicit terms with a factor of $\vec{u}$, the propagator in (\ref{eq:diffeqprop}) also contains an implicit velocity correction arising from the momentum dependence of the potential $v$. For the GW model $v(q)$ depends only on $q^2$ and the particular combination of momenta $(\tilde{p}-\tilde{l})^2$ expanded up to the first subeikonal order reads
\be
(\tilde p - \tilde l)^2 
\approx -(\vec{p}-\vec{l})^2 + \vec{u}\cdot(\vec{p}-\vec{l})
\frac{p_\bot^2-l_\bot^2}{E}\,.
\label{eq:tildediff}
\ee
Without the velocity correction, the $0$-component of $(\tilde{p}-\tilde{l})^\m$ is zero, while the $z$-component of the exchanged momentum can be ignored at leading subeikonal order given that it scales as $\bot^2/E^2$. However, with a non-zero velocity the correction at order $\bot/E$  shown in (\ref{eq:tildediff}) arises, which must be kept since it is of the same order as the other velocity corrections considered up to now. Including only subeikonal contributions at first order, we can write the differential equation satisfied by the in-medium propagator as
\begin{align}
\label{eqG}
\frac{\partial}{\partial L}\mathcal{G}(\vec{p},L; \vec{p}_0,z_0) &= i\left(\vec{u}\cdot\vec{p}-\frac{p_\bot^2}{2E}\right)\mathcal{G}(\vec{p},L; \vec{p}_0,z_0) 
\nonumber\\
&\quad +i\int\frac{d^2\vec{q}}{(2\p)^2}\,\left[1 + \vec{u}\cdot\vec{\Omega}(\vec{p},\vec{q})\right]\,v(q_\bot^2)\,\hat\rho^a(\vec{q},L)\,t^a \,\mathcal{G}(\vec{p}-\vec{q},L; \vec{p}_0,z_0)\, ,
\end{align}
where we have introduced a shorthand notation for the leading subeikonal factors appearing outside of the LPM phases
\be\label{eq:Omega}
\vec{\Omega}(\vec{p},\vec{q}) = -\frac{\vec{p}-\vec{q}}{E}+\frac{\vec{q}}{E}\left(\frac{\left(\vec{p}-\vec{q}\right)^2-\vec{p}^2}{v(q_\bot^2)}\right)\frac{\partial v}{\partial q_\bot^2}\,.
\ee

At this point, all the velocity contributions are explicit. These new terms considerably complicate the differential equation, since now the second term in \eqref{eqG} is not a convolution between the interaction potential and the propagator as in the absence of flow. Working in the eikonal expansion, it is convenient to write the solution of \eqref{eqG} as
\be
{\cal G}(\vec{p},L;\vec{p}_0,z_0) = {\cal G}^{(0)}(\vec{p},L;\vec{p}_0,z_0) + {\cal G}^{(1)}(\vec{p},L;\vec{p}_0,z_0) + {\cal O}\left(\frac{\bot^2}{E^2}\right)\,.\label{eq:expandG}
\ee
Keeping the terms resulting in the LPM phases to all orders, we find that the two contributions ${\cal G}^{(0)}$ and ${\cal G}^{(1)}$ satisfy 
\begin{subequations}
\begin{align}
\frac{\partial}{\partial L}\mathcal{G}^{(0)}(\vec{p},L; \vec{p}_0,z_0) &= i\left(\vec{u} \cdot \vec{p}-\frac{p_{\bot}^2}{2E}\right)\,\mathcal{G}^{(0)}(\vec{p},L; \vec{p}_0,z_0) \nonumber\\
&\quad +i\int\frac{d^2\vec{q}}{(2\p)^2}\,v(q_\bot^2)\,\hat\rho^a(\vec{q},L)t^a\,\mathcal{G}^{(0)}(\vec{p}-\vec{q},L; \vec{p}_0,z_0)\,,
\label{eq:difeqG0} \\
\frac{\partial}{\partial L}\mathcal{G}^{(1)}(\vec{p},L; \vec{p}_0,z_0) &= i\left(\vec{u} \cdot \vec{p}-\frac{p_{\bot}^2}{2E}\right)\mathcal{G}^{(1)}(\vec{p},L; \vec{p}_0,z_0) \nonumber\\
&\quad +i\int\frac{d^2\vec{q}}{(2\p)^2}\,v(q_\bot^2)\,\hat\rho^a(\vec{q},L)t^a\,\mathcal{G}^{(1)}(\vec{p}-\vec{q},L; \vec{p}_0,z_0), \nonumber\\
&\quad +i\int\frac{d^2\vec{q}}{(2\p)^2} \,\vec{u}\cdot\vec{\Omega}(\vec{p},\vec{q})\,v(q_\bot^2)\,\hat\rho^a(\vec{q},L)t^a\,\mathcal{G}^{(0)}(\vec{p}-\vec{q},L; \vec{p}_0,z_0)\,,
\label{eq:difeqG1}
\end{align}
\end{subequations}
with initial conditions
\begin{align}
\mathcal{G}^{(0)}(\vec{p},z_0; \vec{p}_0,z_0) &= (2\pi)^2\delta^{(2)}(\vec{p}-\vec{p}_0)\,, \\
\mathcal{G}^{(1)}(\vec{p},z_0; \vec{p}_0,z_0) &= 0\,.
\end{align}

Let us first look at \eqref{eq:difeqG0}. In the case of static matter this equation has been widely studied, and its solution can be written in terms of a path integral in coordinate space \cite{Casalderrey-Solana:2007knd}. The additional velocity term in this equation introduces only a time-dependent shift in the transverse coordinates of the propagator ${\cal G}^{(0)}(\vec{p},L; \vec{p}_0,z_0)$. In order to see this, we first write \eqref{eq:difeqG0} as
\begin{align}
&\frac{\partial}{\partial L}\left[e^{-i\vec{u}\cdot\vec{p}(L-z_0)}\mathcal{G}^{(0)}(\vec{p},L; \vec{p}_0,z_0)\right] = -i\frac{p_{\bot}^2}{2E}e^{-i\vec{u}\cdot\vec{p}(L-z_0)}\mathcal{G}^{(0)}(\vec{p},L; \vec{p}_0,z_0) \nonumber\\
&\qquad\quad +i\int\frac{d^2\vec{q}}{(2\p)^2}\,e^{-i\vec{u}\cdot\vec{q}(L-z_0)}v(q_\bot^2)\hat\rho^a(\vec{q},L)t^a e^{-i\vec{u}\cdot(\vec{p}-\vec{q})(L-z_0)}\mathcal{G}^{(0)}(\vec{p}-\vec{q},L; \vec{p}_0,z_0)\,,
\end{align}
and notice that the re-scaled function satisfies the same boundary conditions. Taking the Fourier transform into coordinate space, we get
\begin{align}
\label{eq:G0}
\frac{\partial}{\partial L}\mathcal{G}^{(0)}(\vec{x}-(L-z_0)\vec{u},L; \vec{x}_0,z_0) &= \frac{i}{2E}\partial_{\vec{x}}^2\mathcal{G}^{(0)}(\vec{x}-(L-z_0)\vec{u},L; \vec{x}_0,z_0) \nonumber\\
&\hspace{-1cm} +i{\cal A}(\vec{x}-(L-z_0)\vec{u},L)\, \mathcal{G}^{(0)}(\vec{x}-(L-z_0)\vec{u},L; \vec{x}_0,z_0)\,,
\end{align}
where ${\cal A}(\vec{x},L) = \int\frac{d^2\vec{q}}{(2\p)^2}\,e^{i\vec{q}\cdot\vec{x}}\,v(q_\bot^2)\,\hat\rho^a(\vec{q},L)t^a$ is the coordinate space interaction potential weighted with the corresponding color structure. The solution of \eqref{eq:G0} is well known, and can be written as
\begin{align}
\mathcal{G}^{(0)}(\vec{x}-(L-z_0)\vec{u},L; \vec{x}_0,z_0) &= \int\limits_{\vec{x}_0}^{\vec{x}}{\cal D}\vec{r}\exp\left(\frac{iE}{2}\int\limits_{z_0}^L d\xi\,\dot{\vec{r}}^2(\xi)\right)\notag\\
&\hspace{2cm}\times{\cal P}\exp\left(i\int\limits_{z_0}^L d\xi\, {\cal A}(\vec{r}(\xi)-(\xi-z_0)\vec{u},\xi)\right)\,,
\end{align}
or, equivalently,
\begin{align}
\label{G0}
\mathcal{G}^{(0)}(\vec{x},L; \vec{x}_0,z_0) &= \int\limits_{\vec{x}_0}^{\vec{x}+(L-z_0)\vec{u}}{\cal D}\vec{r}\exp\left(\frac{iE}{2}\int\limits_{z_0}^L d\xi\,\dot{\vec{r}}^2(\xi)\right)\notag\\
&\hspace{2cm}\times{\cal P}\exp\left(i\int\limits_{z_0}^L d\xi {\cal A}(\vec{r}(\xi)-(\xi-z_0)\vec{u},\xi)\right)\,.
\end{align}

Now, turning to the subeikonal corrections, one can readily check that the solution of \eqref{eq:difeqG1} can be written in the following form
\begin{align}
\label{G1}
\mathcal{G}^{(1)}(\vec{p},L; \vec{p}_0,z_0) &= i\int_{z_0}^L dz \int\frac{d^2\vec{q}}{(2\pi)^2}\frac{d^2\vec{l}}{(2\pi)^2}\mathcal{G}^{(0)}(\vec{p},L; \vec{l},z)\,(\vec{u}\cdot\vec{\Omega}(\vec{l},\vec{q}))\,v(q_\bot^2)\nonumber\\
&\qquad\times\hat\rho^a(\vec{q},z)t^a \mathcal{G}^{(0)}(\vec{l}-\vec{q},z; \vec{p}_0,z_0)\,.
\end{align}
This expression can also be obtained using the expansion \eqref{eq:expandG} in the integral equation \eqref{eq:propSchwDy} and grouping the terms accordingly.

\section{Resummation at the cross-section level: Broadening}
\label{sec:broadening}
In order to obtain the distribution in transverse momentum, one has to consider the squared amplitude averaged over the scattering centers. From \eqref{eq:amplitudeprop} we get
\be
\label{eq:MMthroughGG}
\left\langle\left|\mathcal{M}(p)\right|^2\right\rangle = \int
\frac{d^2\vec{p}_0}{(2\pi)^2}
\frac{d^2\vec{p}'_0}{(2\pi)^2}
\left\langle J^\dagger(\tilde{p}'_0)\,
\mathcal{G}^\dagger(\vec{p}'_0,0;\vec{p},L)\,
\mathcal{G}(\vec{p},L;\vec{p}_0,0)\,J(\tilde{p}_0)
\right\rangle\,,
\ee
where the arguments of the conjugated propagator are given in reverse order. 

The medium averaging acts only on the factors of $\hat\rho^a$ appearing in the propagators. For simplicity, we take a Gaussian distribution for the scatterings in which the only non-trivial correlation is the two-point function
\be
\label{eq:bgfield}
\left\langle\hat\rho^a(\vec{x}, z_x)\hat\rho^b(\vec{y}, z_y)\right\rangle = \frac{1}{2C_{\bar R}}\delta^{ab}\,\delta^{(2)}(\vec{x}-\vec{y})\,\delta(z_x-z_y)\,\rho(\vec{x}, z_x)\,,
\ee
where we have defined $C_{\bar R}$ as $C_{\bar R}=N_c$ for quarks and $C_{\bar R}=C_F$ for gluons. We have also introduced  the number density of scattering centers $\rho(\vec{x}, z)$. This averaging procedure coincides with the usual GW model used in \cite{Sadofyev:2021ohn,Barata:2022krd} as well as the McLerran-Venugopalan model in the CGC context. Taking the transverse Fourier transform of \eqref{eq:bgfield} we get
\be
\left\langle\hat\rho^a(\vec{q},z)\hat\rho^b(\vec{q}',z')\right\rangle 
= \frac{1}{2C_{\bar R}}\delta^{ab}\delta(z-z')
\int d^2\vec{x}\, e^{-i(\vec{q}+\vec{q}')\cdot\vec{x}}\,\rho(\vec{x},z)\,.
\label{eq:FTaverage}
\ee

From now on, we will focus on the case of transversely homogeneous matter, setting $\rho$ to be a function of $z$ only. The leading effects of  medium transverse anisotropies on jet-medium interactions, manifested through gradients of the thermodynamic parameters, were already discussed in \cite{Sadofyev:2021ohn,Barata:2022krd}. In fact, in the absence of transverse flow, the gradient contributions to jet broadening are also subeikonal (although enhanced by the medium length), and can be clearly separated from the flow effects up to first subeikonal order. Cross terms involving both gradients and transverse velocity have not yet been addressed in the literature, and we leave them for future work. It is worth noticing, however, that the leading cross contributions affect only the even moments of the final distribution. 

For transversely homogeneous matter $\rho(\vec{x},z)=\rho(z)$, and we can perform the Fourier transformation in \eqref{eq:FTaverage} getting 
\be
\label{eq:avrhomomentum}
\left\langle\hat\rho^a(\vec{q},z)\hat\rho^b(\vec{q}',z')\right\rangle 
= \frac{1}{2C_{\bar R}} \delta^{ab}\delta(z-z')\delta^{(2)}(\vec{q}+\vec{q}')\,\rho(z)\,.
\ee
Note that since $\hat\rho^a$ is real in coordinate space, $\hat\rho^{a\dagger}(\vec{q}_\bot,z) = \hat\rho^a(-\vec{q}_\bot,z)$. 

The source terms $J$ and $J^\dagger$ can be pulled out of the medium averages, allowing us to focus on the combination of two propagators appearing in \eqref{eq:MMthroughGG}. Moreover, the trivial color structure of the averaging procedure \eqref{eq:bgfield} implies
\be
\left\langle \mathcal{G}^\dagger(\vec{p}'_0,0;\vec{p},L)\,
\mathcal{G}(\vec{p},L;\vec{p}_0,0)\right\rangle 
= 
\frac{\bf 1}{d_{proj}}\left\langle\text{Tr}\left[\mathcal{G}^\dagger(\vec{p}'_0,0;\vec{p},L)\,
\mathcal{G}(\vec{p},L;\vec{p}_0,0)\right]\right\rangle\,,
\label{eq:GGtrace}
\ee
where $d_{proj}$ is the dimension of the representation corresponding to the projectile.

One could in principle calculate this average of two propagators directly using their explicit expressions~\eqref{G0}~and~\eqref{G1}, but that would entail summing all the contributions from all possible pairings of the explicit factor of $\hat\rho$ appearing in \eqref{G1} with other $\hat\rho$'s hidden in the factors of $\mathcal{G}^{(0)}$. Instead, here we derive a differential equation in $L$ satisfied by this product of propagators, similarly to \cite{Isaksen:2020npj}. We start by considering an infinitesimal step in $L$ and using the convolution properties of the propagators we have
\begin{align}
\label{eq:GG}
&\frac{1}{d_{proj}}
\left\langle\text{Tr}\left[\mathcal{G}^\dagger(\vec{p}'_0,0;\vec{p},L+\epsilon)\,
\mathcal{G}(\vec{p},L+\epsilon;\vec{p}_0,0)\right]\right\rangle 
\nonumber\\
&= 
\int_{\vec{l},\vec{l}'}\frac{1}{d_{proj}}
\left\langle\text{Tr}\left[\mathcal{G}^\dagger(\vec{p}'_0,0;\vec{l}',L)\,
\mathcal{G}^\dagger(\vec{l}',L;\vec{p},L+\epsilon)\,
\mathcal{G}(\vec{p},L+\epsilon;\vec{l},L)\,
\mathcal{G}(\vec{l},L;\vec{p}_0,0)\right]\right\rangle\,.
\end{align}
From \eqref{eq:avrhomomentum} it is clear that the averages are local in the longitudinal coordinate, meaning that only pairs of $\hat\rho$'s at the same longitudinal coordinate give a non-zero contribution to the average. This allows us to factorize the average in \eqref{eq:GG} as the product of averages of propagators having a common support in the longitudinal direction. Combining this fact with the color triviality, as seen for instance in \eqref{eq:GGtrace}, we can re-write
the trace in the r.h.s. of \eqref{eq:GG} as
\begin{align}
&\frac{1}{d_{proj}}
\left\langle\text{Tr}\left[\mathcal{G}^\dagger(\vec{p}'_0,0;\vec{p},L+\epsilon)\,\mathcal{G}(\vec{p},L+\epsilon;\vec{p}_0,0)\right]\right\rangle 
\nonumber\\
&\hspace{0cm}= 
\int_{\vec{l},\vec{l}'}\frac{1}{d_{proj}}
\left\langle\text{Tr}\left[\mathcal{G}^\dagger(\vec{l}',L;\vec{p},L+\epsilon)\mathcal{G}(\vec{p},L+\epsilon;\vec{l},L)\right]\right\rangle\frac{1}{d_{proj}}
\left\langle\text{Tr}\left[\mathcal{G}^\dagger(\vec{p}'_0,0;\vec{l}',L)\mathcal{G}(\vec{l},L;\vec{p}_0,0)\right]\right\rangle\,.\notag
\end{align}
In order to find the terms linear in $\epsilon$, one has to expand the propagators going from $L$ to $L+\epsilon$ using iteratively the integral relation \eqref{eq:propSchwDy}. At the required accuracy, we have
\begin{align}
&\mathcal{G}(\vec{p},L+\epsilon;\vec{l},L) \simeq (2\pi)^2\delta^{(2)}(\vec{p}-\vec{l})\,e^{i\left(\vec{u}\cdot\vec{p}-\frac{p_\bot^2}{2E}\right)\epsilon}\nonumber\\
&\qquad
+i\int_L^{L+\epsilon}dz\int\frac{d^2\vec{q}}{(2\pi)^2}\left(1+\vec{u}\cdot\vec{\Omega}(\vec{p},\vec{q})\right)\,v(q_\bot^2)\,\hat\rho^a(\vec{q},z)\,t^a\,e^{i\left(\vec{u}\cdot\vec{p}-\frac{p_\bot^2}{2E}\right)(L+\epsilon-z)}
\nonumber\\
&\qquad\quad\times\,e^{i\left(\vec{u}\cdot\vec{l}-\frac{l_\bot^2}{2E}\right)(z-L)}(2\pi)^2\delta^{(2)}(\vec{p}-\vec{q}-\vec{l})
\nonumber\\
&\qquad -\int_L^{L+\epsilon}dz_2\int_L^{z_2}dz_1\int\frac{d^2\vec{q}_2}{(2\pi)^2}\frac{d^2\vec{q}_1}{(2\pi)^2}\left(1+\vec{u}\cdot\vec{\Omega}(\vec{p},\vec{q}_2)\right)\left(1+\vec{u}\cdot\vec{\Omega}(\vec{p}-\vec{q}_2,\vec{q}_1)\right)
\nonumber\\
&\qquad\quad\times v(q_{2\bot}^2)\,v(q_{1\bot}^2)\,\hat\rho^{a_2}(\vec{q}_2,z_2)\,t^{a_2}\,\hat\rho^{a_1}(\vec{q}_1,z_1)\,t^{a_1}\,e^{i\left(\vec{u}\cdot\vec{p}-\frac{p_\bot^2}{2E}\right)(L+\epsilon-z_2)}e^{i\left(\vec{u}\cdot(\vec{p}-\vec{q}_2)-\frac{(\vec{p}-\vec{q}_2)^2}{2E}\right)(z_2-z_1)}
\nonumber\\
&\qquad\quad\times\,e^{i\left(\vec{u}\cdot\vec{l}-\frac{l_\bot^2}{2E}\right)(z_1-L)}(2\pi)^2\delta^{(2)}(\vec{p}-\vec{q}_2-\vec{q}_1-\vec{l})\,,
\label{eq:exppropeps}
\end{align}
with $\mathcal{G}^\dagger(\vec{l}',L;\vec{p},L+\epsilon)$ following from it.

When we take the average of the trace of two propagators, the linear terms in $\hat\rho$ cancel out, while the terms with exactly two factors of $\hat\rho$ yield to the linear contribution in $\epsilon$. The latter
come either from the product of a term with no $\hat\rho$'s in one of the propagators times a term with two $\hat\rho$'s in the other one, the so-called double Born contributions, or from the product of two terms with one $\hat\rho$ each, referred to as single Born contributions.

Let us first consider the double Born contributions, involving the last term in~\eqref{eq:exppropeps} (or its conjugate). According to \eqref{eq:avrhomomentum}, we have an additional $\delta$-function setting $\vec{q}_1=-\vec{q}_2$, which also ensures that all the phases cancel out. The corresponding contribution takes the form
\begin{align}
\label{SBsimplified}
&-\mathcal{C}\int_L^{L+\epsilon}dz_2\int_L^{z_2}dz_1\int\frac{d^2\vec{q}}{(2\pi)^2}\left(1+\vec{u}\cdot\vec{\Omega}(\vec{p},\vec{q})\right)\left(1+\vec{u}\cdot\vec{\Omega}(\vec{p}-\vec{q},-\vec{q})\right)
\nonumber\\
&\qquad\quad\times v^2(q_\bot^2)\,\delta(z_2-z_1)\,\rho(z_1)\,(2\pi)^2\delta^{(2)}(\vec{p}-\vec{l})(2\pi)^2\delta^{(2)}(\vec{p}-\vec{l}')\,,
\end{align}
where we have combined the color structure coming from the projectile generators with the color factor in \eqref{eq:avrhomomentum}, introducing $\mathcal{C}=\frac{C_{proj}}{2C_{\bar{R}}}$, where $C_{proj}$ is the Casimir of the projectile representation. One may notice that the double $z$-integration in (\ref{SBsimplified}) is not fully defined, since it involves a delta function on the edge of the domain. However, due to the locality of the medium average \eqref{eq:avrhomomentum}, the result has to be symmetric under the exchange of $z_2$ and $z_1$ in \eqref{SBsimplified}, and thus this particular $\d$-function is symmetric, yielding
\be
\int_L^{L+\epsilon}dz_2\int_L^{z_2}dz_1\,\delta(z_2-z_1)\rho(z_1) 
\simeq \frac{\epsilon}{2}\rho(L)\,.
\ee
Keeping only the leading subeikonal corrections, we can write
\be
\left(1+\vec{u}\cdot\vec{\Omega}(\vec{p},\vec{q})\right)\left(1+\vec{u}\cdot\vec{\Omega}(\vec{p}-\vec{q},-\vec{q})\right)
\simeq 1 + \vec{u} \cdot\left[\vec{\Omega}(\vec{p},\vec{q})+\vec{\Omega}(\vec{p}-\vec{q},-\vec{q})\right]\,,
\ee
and thus, the $\vec{q}$-integration in \eqref{SBsimplified} reduces to
\begin{align}
&\int\frac{d^2\vec{q}}{(2\pi)^2}\left(1 + \vec{u} \cdot\left[\vec{\Omega}(\vec{p},\vec{q})+\vec{\Omega}(\vec{p}-\vec{q},-\vec{q})\right]\right)v^2(q_\bot^2) \nonumber\\
&=\int\frac{d^2\vec{q}}{(2\pi)^2}\left(1 + 2\,\vec{u}\cdot\vec{\Omega}(\vec{p},\vec{q})\right)v^2(q_\bot^2)\,,
\end{align}
where we have used the explicit form of $\vec{\Omega}$ in \eqref{eq:Omega}.

Now, let us consider the term in which each propagator contributes with a factor of $\hat\rho$. The additional $\delta$-function coming from the average of the $\hat\rho$'s sets the momentum exchange equal on both the amplitude and conjugate amplitude. Then, this contribution takes the following form
\begin{align}
&\mathcal{C}\int_L^{L+\epsilon}dz
\int_L^{L+\epsilon}dz'
\int\frac{d^2\vec{q}}{(2\pi)^2}
\left(1+\vec{u}\cdot\vec{\Omega}(\vec{p},\vec{q})\right)^2v^2(q_\bot^2)\delta(z-z')\rho(z)
\nonumber\\
&\qquad\times\,(2\pi)^2\delta^{(2)}(\vec{p}-\vec{q}-\vec{l})(2\pi)^2\delta^{(2)}(\vec{p}-\vec{q}-\vec{l}')\,.
\end{align}
Since the line $z-z'=0$ is always within the integration domain, the $z$-integrations can be readily performed and one finds
\be
\int_L^{L+\epsilon}dz\int_L^{L+\epsilon}dz'\,\delta(z-z')\rho(z) \simeq \epsilon\rho(L)\,.
\ee
Up to leading subeikonal order, we also have
\be
\left(1+\vec{u}\cdot\vec{\Omega}(\vec{p},\vec{q})\right)^2
\simeq 1 + 
2\,\vec{u}\cdot\vec{\Omega}(\vec{p},\vec{q})\,,
\ee
and putting the single and the two double Born contributions together it is easy to see that
\begin{align}
\label{GGepsilon}
&\frac{1}{d_{proj}}
\left\langle\text{Tr}\left[\mathcal{G}^\dagger(\vec{l}',L;\vec{p},L+\epsilon)\,\mathcal{G}(\vec{p},L+\epsilon;\vec{l},L)\right]\right\rangle\ 
= (2\pi)^4\delta^{(2)}(\vec{p}-\vec{l})\delta^{(2)}(\vec{l}-\vec{l}')
\nonumber\\
&\qquad\qquad - \epsilon\int\frac{d^2\vec{q}}{(2\pi)^2}\,\sigma(\vec{p},\vec{q}; L)\,(2\pi)^4\delta^{(2)}(\vec{p}-\vec{q}-\vec{l})\delta^{(2)}(\vec{l}-\vec{l}') + \mathcal{O}(\epsilon^2)\,,
\end{align}
where we have introduced $\sigma(\vec{p},\vec{q};L)$, a specific combination of the in-medium color potentials in the GW model given by
\begin{align}
\sigma(\vec{p},\vec{q}; L) &= -\mathcal{C}\,\rho(L)\,
\left(1+2\,\vec{u}\cdot\vec{\Omega}(\vec{p},\vec{q})\right)\,v^2(q_\bot^2) \notag\\
&\hspace{3cm}+ \mathcal{C}\,\rho(L)\,\delta^{(2)}(\vec{q})
\int\,d^2\vec{l}\,\left(1+2\,\vec{u}\cdot\vec{\Omega}(\vec{p},\vec{l})\right)\,v^2(l_\bot^2)\,.
\end{align}
Its eikonal limit is related to the forward scattering amplitude for a color dipole, and it is often referred to as the dipole cross-section or dipole potential. 

Eq.~\eqref{GGepsilon} gives both the $L$-evolution and the initial condition of the average of two propagators. Indeed, taking both $L=0$ and $\epsilon=0$ in \eqref{GGepsilon}, the initial condition reads
\be
\frac{1}{d_{proj}}
\left\langle\text{Tr}\left[\mathcal{G}^\dagger(\vec{p}'_0,0;\vec{p},0)\,
\mathcal{G}(\vec{p},0;\vec{p}_0,0)\right]\right\rangle=(2\p)^4\d^{(2)}(\vec{p}-\vec{p}_0)\d^{(2)}(\vec{p}_0-\vec{p}'_0)\,,
\label{eq:init_cond}
\ee
It is worth noticing that the $L$-evolution in \eqref{GGepsilon} is given by a convolution, involving only the final momenta, and thus the delta function forcing the initial momenta to be equal in the initial condition \eqref{eq:init_cond} is still present for arbitrary $L$, and we can write
\be
\frac{1}{d_{proj}}
\left\langle\text{Tr}\left[\mathcal{G}^\dagger(\vec{p}'_{0},0;\vec{p}_,L)\,
\mathcal{G}(\vec{p},L;\vec{p}_{0},0)\right]\right\rangle
\equiv 
(2\pi)^2\delta^{(2)}(\vec{p}_{0}-\vec{p}'_{0})\,\mathcal{P}(\vec{p},L;\vec{p}_{0},0)\,,\label{eq:defP}
\ee
where $\mathcal{P}(\vec{p},L;\vec{p}_{0},0)$ is the broadening probability. The differential equation on the average of two propagators $\frac{1}{d_{proj}}
\left\langle\text{Tr}\left[\mathcal{G}^\dagger(\vec{p}'_0,0;\vec{p},L)
\,\mathcal{G}(\vec{p},L;\vec{p}_0,0)\right]\right\rangle$ can now be reduced to
\begin{align}
\label{eqP}
\frac{\partial}{\partial L}\mathcal{P}(\vec{p},L;\vec{p}_{0},0)
= - \int\frac{d^2\vec{q}}{(2\pi)^2}\,\sigma(\vec{p},\vec{q}; L)
\,\mathcal{P}(\vec{p}-\vec{q},L;\vec{p}_{0},0)\,,
\end{align}
with initial condition
\be
\mathcal{P}(\vec{p},0;\vec{p}_{0},0) 
= (2\pi)^2\delta^{(2)}(\vec{p}-\vec{p}_{0})\,.
\ee

We can further consider this differential equation order by order in inverse powers of $E$. At zeroth order we have
\be
\frac{\partial}{\partial L}\mathcal{P}^{(0)}(\vec{p},L;\vec{p}_{0},0)
= 
-\int\frac{d^2\vec{q}}{(2\pi)^2}\,\mathcal{V}(\vec{q};L)\,\mathcal{P}^{(0)}(\vec{p}-\vec{q},L;\vec{p}_0,0)\,,
\label{eq:P0_eik}
\ee
with initial condition
\be
\mathcal{P}^{(0)}(\vec{p},0;\vec{p}_{0},0) 
= (2\pi)^2\delta^{(2)}(\vec{p}-\vec{p}_{0})\,,
\ee
where the superscript corresponds to the order in the eikonal expansion, and we have defined $\mathcal{V}$, the eikonal limit of the dipole cross-section 
\be
\label{eq:sigma_eikonal}
\mathcal{V}(\vec{q};L)\equiv
\sigma(\vec{p},\vec{q}; L)|_{\vec{u}=0}\,.
\ee

Eq.~(\ref{eq:P0_eik}) can be easily solved in coordinate space. Focusing for simplicity on the case of constant $\rho$, we find, as expected, the standard result for the broadening probability in static uniform matter
\be
\label{eq:Pzero}
\mathcal{P}^{(0)}(\vec{r},L;\vec{r}_{0},0)
= 
e^{-\mathcal{V}(\vec{r})L}\delta^{(2)}(\vec{r}-\vec{r}_{0})\,,
\ee
which in momentum space reads
\be
\mathcal{P}^{(0)}(\vec{p},L;\vec{p}_{0},0)
= \int d^2\vec{r} 
\,e^{-i(\vec p-\vec p_0)\vec r}\, e^{-\mathcal{V}(\vec{r})L}\,.
\ee

Using this solution, we can obtain the leading subeikonal contribution to the differential equation \eqref{eqP}, which reads
\begin{align}
\label{eqP1}
\frac{\partial}{\partial L}\mathcal{P}^{(1)}(\vec{p},L;\vec{p}_{0},0) 
&= 
-\int\frac{d^2\vec{q}}{(2\pi)^2}\,
\mathcal{V}(\vec{q})\,\mathcal{P}^{(1)}(\vec{p}-\vec{q},L;\vec{p}_{0},0)
\nonumber\\
&\quad-\int\frac{d^2\vec{q}}{(2\pi)^2}\,\left[\sigma(\vec{p},\vec{q})-\mathcal{V}(\vec{q})\right]\,\mathcal{P}^{(0)}(\vec{p}-\vec{q},L;\vec{p}_{0},0)\,.
\end{align}

Let us take a closer look at the subeikonal contribution to the interaction potential, we can re-express the corresponding kernel as
\begin{align}
\sigma(\vec{p},\vec{q})-\mathcal{V}(\vec{q})
&= -2\frac{\vec{u}\cdot(\vec{p}-\vec{q})}{E}\mathcal{V}(\vec{q}) - \frac{u_\a}{E}\,\mathcal{V}_{\a\b}(\vec{q})\left(2(\vec{p}-\vec{q}
)_\b + \vec{q}_\b\right)\,,
\end{align}
with
\begin{align}
\mathcal{V}_{\a\b}(\vec{q})&= \mathcal{C}\,\r\,\left[-\vec{q}_\a\vec{q}_\b\frac{\partial v^2}{\partial q_\bot^2} - (2\pi)^2\delta^{(2)}(\vec{q})\frac{\delta_{\a\b}}{2}\int\frac{d^2\vec{l}}{(2\pi)^2}\,v^2(l_\bot^2)\right]\,,\label{eq:Vab}
\end{align}
where $\a$ and $\b$ run over the transverse 2D subspace. Taking the Fourier transform of the convolution in \eqref{eqP1}, we can replace the factors of $\vec{p}-\vec{q}$ by derivatives acting on $\mathcal{P}^{(0)}$ and the factors of $\vec{q}$ alone by derivatives acting on $v^2(q_\bot^2)$. Then, the differential equation (\ref{eqP1}) can be written in the following compact form
\begin{align}
\label{eqP1x}
\frac{\partial}{\partial L}\mathcal{P}^{(1)}(\vec{r},L;\vec{r}_0,0) 
&= -\mathcal{V}(\vec{r})\,\mathcal{P}^{(1)}(\vec{r},L;\vec{r}_0,0) \nonumber\\
&\quad - i\frac{u_\a}{E}\,\Big[2\Big(\mathcal{V}(\vec{r})\delta_{\alpha\beta} + \mathcal{V}_{\a\b}(\vec{r})\Big)\na_\b + \na_\b\mathcal{V}_{\a\b}(\vec{r})\Big]\mathcal{P}^{(0)}(\vec{r},L;\vec{r}_0,0)\,,
\end{align}
where ${\cal V}_{\alpha\beta}(\vec{r})$ is the Fourier transform of ${\cal V}_{\alpha\beta}(\vec{q})$, which can be written in terms of ${\cal V}(\vec{r})$ as
\be 
{\cal V}_{\alpha\beta}(\vec{r}) = -\frac{\delta_{\alpha\beta}}{2}{\cal V}(\vec{r}) - \vec{r}_\alpha\vec{r}_\beta \frac{\partial}{\partial r_\bot^2}{\cal V}(\vec{r}).
\ee
Plugging now \eqref{eq:Pzero} into the equation above, the  integration over $L$ can be easily performed. Using the trivial initial condition $\mathcal{P}^{(1)}(\vec{r},0;\vec{r}_0,0)=0$, we find
\begin{align}
\mathcal{P}^{(1)}(\vec{r},L;\vec{r}_0,0) &= e^{-\mathcal{V}(\vec{r})L}\,\frac{u_\a}{E}\Big\{-2iL\,\Big(\mathcal{V}(\vec{r})\delta_{\a\b} + \mathcal{V}_{\a\b}(\vec{r})\Big)\na_\b\,\delta^{(2)}(\vec{r}-\vec{r}_0)
\nonumber\\
&\quad+\Big[-i L\,\na_\b\mathcal{V}_{\a\b}(\vec{r})+iL^2\Big(\mathcal{V}(\vec{r})\delta_{\a\b} + \mathcal{V}_{\a\b}(\vec{r})\Big)\na_\b\mathcal{V}(\vec{r})\Big]\delta^{(2)}(\vec{r}-\vec{r}_0)\Big\}\,.\label{eq:P1coor}
\end{align}
Finally, taking the Fourier transform back to momentum space, we have
\begin{align}
\mathcal{P}^{(1)}(\vec{p},L;\vec{p}_0,0) &= 
\int d^2\vec{r} \,e^{-i(\vec{p}-\vec{p}_0)\cdot\vec{r}}e^{-\mathcal{V}(\vec{r})L}\,\frac{u_\a}{E}\,
\Big[2L\,\vec{p}_\b\,\Big(\mathcal{V}(\vec{r})\delta_{\a\b} + \mathcal{V}_{\a\b}(\vec{r})\Big) \nonumber\\
& \quad+i L\,\na_\b\mathcal{V}_{\a\b}(\vec{r})+2i L\,\na_{\a}\mathcal{V}(\vec{r})-iL^2\,
\Big(\mathcal{V}(\vec{r})\delta_{\a\b} + \mathcal{V}_{\a\b}(\vec{r})\Big)\na_\b\mathcal{V}(\vec{r})
\Big]\,,
\end{align}
or, alternatively, after some algebra and integrating by parts
\begin{align}
\mathcal{P}^{(1)}(\vec{p},L;\vec{p}_0,0) &= 
\int d^2\vec{r}\, e^{-i(\vec{p}-\vec{p}_0)\cdot\vec{r}}e^{- \mathcal{V}(\vec{r})L}\,\frac{u_\a}{E}\,
\Big[2L\,\vec{p}_{0\b}\,\Big(\mathcal{V}(\vec{r})\delta_{\a\b} + \mathcal{V}_{\a\b}(\vec{r})\Big) \nonumber\\
&\quad-i L\,\na_\b\mathcal{V}_{\a\b}(\vec{r})
+iL^2\,\Big(\mathcal{V}(\vec{r})\delta_{\a\b} + \mathcal{V}_{\a\b}(\vec{r})\Big)\na_\b\mathcal{V}(\vec{r})
\Big]\,.\label{eq:P1withp0}
\end{align}
This leading subeikonal correction to the broadening probability distribution\footnote{Notice that the momentum space form of the distribution is in fact real as expected from the form of \eqref{eqP1}.} resulting from the medium flow effects is one of the main results of this paper. For completeness, we show in the appendix~\ref{ap:A} the results for the eikonal and leading subeikonal contributions to the broadening in the so-called harmonic oscillator regime, in which the dipole cross-section is approximated by its leading logarithmic behavior ${\cal V}(\vec{r}) = \hat q r_\bot^2/4$.

We can now use Eqs.~\eqref{eq:MMthroughGG},~\eqref{eq:GGtrace},~and~\eqref{eq:defP} to write the amplitude squared as
\be
\left\langle\left|\mathcal{M}(p)\right|^2\right\rangle = \int
\frac{d^2\vec{p}_0}{(2\pi)^2}\,\mathcal{P}(\vec{p},L;\vec{p}_0,0)\left|J(\tilde p_0)\right|^2\, .
\ee
Recognizing that the leading subeikonal expansion of the source term enters as a shift in the energy, as explained in \cite{Sadofyev:2021ohn}, we get
\be
|J(\tilde p_0)|^2 = |J(E,\vec{p}_0)|^2 - \vec{u}\cdot(\vec{p}-\vec{p}_0)\frac{\partial}{\partial E}|J(E,\vec{p}_0)|^2\,.
\ee
Then, at leading subeikonal order, the amplitude squared takes the form
\begin{align}
\left\langle\left|\mathcal{M}(p)\right|^2\right\rangle &= \int
\frac{d^2\vec{p}_0}{(2\pi)^2}\,\bigg[\mathcal{P}^{(0)}(\vec{p},L;\vec{p}_0,0)\nonumber\\
&\quad+ \left.\left(\mathcal{P}^{(1)}(\vec{p},L;\vec{p}_0,0)-\vec{u}\cdot(\vec{p}-\vec{p}_0)\mathcal{P}^{(0)}(\vec{p},L;\vec{p}_0,0)\frac{\partial}{\partial E}\right)\right]|J(E,\vec{p}_0)|^2\,.\label{eq:expandedM2}
\end{align}

\section{Final-state distribution and its moments}
\label{sec:moments}
The final state jet momentum distribution can be written in terms of the squared scattering amplitude as
\bea
E\frac{d\N}{d^2\tvec{p}\,dE}\equiv\frac{1}{2(2\pi)^3}\,\left\langle \left| \mathcal{M}(p) \right|^2 \right\rangle\,,
\eea
which, using Eq.~\eqref{eq:expandedM2}, can be related to the initial state distribution $E\frac{d\N^{(0)}}{d^2\tvec{p}_0dE}\equiv\frac{1}{2(2\pi)^3}|J(E,\vec{p}_0)|^2$ by
\begin{align}
E\frac{d\N}{d^2\tvec{p}\,dE} &= \int
\frac{d^2\vec{p}_0}{(2\pi)^2}\,\bigg[\mathcal{P}^{(0)}(\vec{p},L;\vec{p}_0,0)\nonumber\\
&\quad+ \left.\left(\mathcal{P}^{(1)}(\vec{p},L;\vec{p}_0,0)-\vec{u}\cdot(\vec{p}-\vec{p}_0)\mathcal{P}^{(0)}(\vec{p},L;\vec{p}_0,0)\frac{\partial}{\partial E}\right)\right]E\frac{d\N^{(0)}}{d^2\tvec{p}_0\,dE}\,.
\label{eq:momdist}
\end{align}

It is well known that in the exact eikonal limit the final distribution can be written as a convolution of the medium effects and the initial distribution, which implies a factorization in coordinate space. However, the leading subeikonal terms related to the medium flow clearly break this separation, since the final distribution is sensitive to the slope of the energy dependence of the initial distribution, as it can be clearly seen in (\ref{eq:momdist}). This situation is similar to the case of broadening in inhomogeneous matter \cite{Barata:2022krd}, where a spatial gradient of the coordinate space representation of the initial distribution appears coupled to the transverse gradients of the hydrodynamic parameters of the matter at leading subeikonal order, thus breaking the factorization. The main difference between these two types of leading subeikonal corrections relies in the fact that the energy dependence of the initial distribution plays a crucial role in the description of most jet observables, while this is not the case for its transverse spatial gradients. Hence, one may model the initial distribution in a way that its transverse spatial gradients vanish, taking the limit of a narrow distribution in momentum space $E\frac{d\N^{(0)}}{d^2\vec{p}dE}\sim\d^{(2)}(\vec{p})$ as done in \cite{Barata:2022krd}, whereas one can never disregard the energy derivative of the initial distribution in Eq.~(\ref{eq:momdist}).

Given that the derivation of (\ref{eq:momdist}) is kinematically restricted to $t$-channel interactions, the number of jets (partons) cannot change due to medium effects
\bea
\int\,d^2\tvec{p}\,\frac{d\N}{d^2\tvec{p}\,dE}=\int\,d^2\tvec{p}\,\frac{d\N^{(0)}}{d^2\tvec{p}\,dE}\,.
\eea
Since this unitarity condition is clearly satisfied by the eikonal broadening, we only need to check it for the flow effects arising at leading subeikonal order
\begin{align}
\label{unitarity}
\int
\frac{d^2\vec{p}\,d^2\vec{p}_0}{(2\pi)^4}\, \left(\mathcal{P}^{(1)}(\tvec{p},L;\tvec{p}_0,0)-\tvec{u}\cdot(\tvec{p}-\tvec{p}_0)\,\mathcal{P}^{(0)}(\tvec{p},L;\tvec{p}_0,0)\frac{\pa}{\pa E}\right)\,E\frac{d\N^{(0)}}{d^2\tvec{p}_0\,dE}=0\,.
\end{align}
First we notice that the angular integration in the second term in (\ref{unitarity}) is zero since $\mathcal{P}^{(0)}(\vec{p},L;\vec{p}_{0},0)$ is a function of $|\vec{p}-\vec{p}_{0}|$ only, as can be seen from \eqref{eq:Pzero}. For the first term in (\ref{unitarity}), the $\vec p$-integration results in a delta function setting $\tvec r=0$ in the integrand of (\ref{eq:P1withp0}). One can readily check that $\mathcal{V}(\tvec 0)=0$ and $\mathcal{V}_{\a\b}(\tvec 0)=0$, as well as their first gradients, thus, giving a vanishing contribution which ensures unitarity.

It was shown in \cite{Sadofyev:2021ohn} that, at leading order in opacity, flow effects lead to non-zero odd moments of the final momentum distribution, which would vanish in the case of isotropic broadening when the flow velocity is zero. Here, we extend this result to the case of multiple scatterings, accurate to all orders in the opacity expansion. To compare to the results in \cite{Sadofyev:2021ohn}, let us consider the same family of moments of the momentum distribution
\begin{align}
\left\langle \tvec{F}(\tvec p)\right\rangle=
\frac{\int\,d^2\tvec p\,\tvec{F}(\tvec p)\,\frac{d\N}{d^2\tvec{p}dE}}{\int\,d^2\tvec p\,\frac{d\N^{(0)}}{d^2\tvec{p}dE}}\,,
\label{eq:momentdist}
\end{align}
with $\tvec{F}(\tvec p)=p_\bot^{2k} \tvec{p}$, and assume that the initial distribution is highly collimated
\begin{align}
\label{N0delta}
    E \frac{dN^{(0)}}{d^2p\, dE} = f(E) \, \delta^{(2)} (\tvec{p}) \,,
\end{align}
where $f(E)$ is some arbitrary energy dependence.

We choose to focus on the odd moments since they are zero in the eikonal limit but receive a correction at first subeikonal order, as opposed to the even moments which are non-zero in the eikonal limit but the first non-trivial correction appears at second subeikonal order. Focusing on the first term of (\ref{eq:momdist}) and its contributions to (\ref{eq:momentdist}), it can be easily seen why in the eikonal limit the odd moments of the momentum distribution are zero. Indeed,
\begin{align}
\left\langle p_\bot^{2k} \tvec{p} \right\rangle^{(0)}=\int\frac{d^2\vec{p}\,d^2\vec{r}}{(2\pi)^2}\,
p_\bot^{2k} \tvec{p}\,e^{-i\vec{p}\cdot\vec{r}}e^{-\mathcal{V}(\vec{r})L}=0
\,,
\end{align}
since $\mathcal{V}(\vec{r})$ depends only on the magnitude $r_\bot$ and therefore the two angular integrations vanish. Here and in what follows, the superscript of averages indicates the particular order of the expressions in eikonal expansion. The correction to these moments at first subeikonal order is given by
\begin{align}
\label{AveragedP}
\left\langle p_\bot^{2k} \tvec{p} \right\rangle^{(1)}&=\int\frac{d^2\vec{p}\,d^2\vec{r}}{(2\pi)^2}\, p_\bot^{2k} \tvec{p}\,e^{-i\vec{p}\cdot\vec{r}}e^{-\mathcal{V}(\vec{r})L}\,\frac{u_\a}{E}
\notag\\
&\hspace{1cm}\times
\Bigg[iL^2\,
\Big(\mathcal{V}(\vec{r})\delta_{\a\b} + \mathcal{V}_{\a\b}(\vec{r})\Big)
\na_\b\mathcal{V}(\vec{r})-
iL\,\na_\b\mathcal{V}_{\a\b}(\vec{r})
-E\frac{f'(E)}{f(E)}\,p_\a
\Bigg]
\notag\\
&=-\frac{\vec{u}}{2}\,\frac{f'(E)}{f(E)}\left\langle p_\bot^{2k+2} \right\rangle+\int\frac{d^2\vec{p}\,d^2\vec{r}}{(2\pi)^2}\, p_\bot^{2k} \tvec{p}\,e^{-i\vec{p}\cdot\vec{r}}e^{-\mathcal{V}(\vec{r})L}\,\frac{u_\a}{E}
\notag\\
&\hspace{1cm}\times\Bigg[iL^2\,
\Big(\mathcal{V}(\vec{r})\delta_{\a\b} + \mathcal{V}_{\a\b}(\vec{r})\Big)\na_\b\mathcal{V}(\vec{r})
-iL\,\na_\b\mathcal{V}_{\a\b}(\vec{r})
\Bigg]
\,,
\end{align}
where the even momentum $\left\langle p_\bot^{2k+2} \right\rangle$ is calculated in the eikonal limit, since its first non-vanishing subeikonal corrections appear at second subeikonal order and $f(E)$ scales as $E^{-n}$ with $n>0$ (and thus the ratio $f'(E)/f(E) = -n/E$). Replacing the momentum vector $\vec{p}$ by the spatial gradient in (\ref{AveragedP}), integrating by parts, and averaging over the angles in the spatial integral we get
\begin{align}
\left\langle p_\bot^{2k} \vec{p} \right\rangle^{(1)}
&=-\frac{\vec{u}}{2}\,\frac{f'(E)}{f(E)}\left\langle p_\bot^{2k+2} \right\rangle-i\frac{\vec{u}}{2E}\int\frac{d^2\vec{p}\,d^2\vec{r}}{(2\pi)^2}\, p_\bot^{2k}\,e^{-i\vec{p}\cdot\vec{r}}
\notag\\
&\hspace{1cm}
\times \na_\a
\Bigg\{e^{-\mathcal{V}(\vec{r})L}
\Bigg[iL^2\,
\Big(\mathcal{V}(\vec{r})\delta_{\a\b} + \mathcal{V}_{\a\b}(\vec{r})\Big)\na_\b\mathcal{V}(\vec{r})
-iL\,\na_\b\mathcal{V}_{\a\b}(\vec{r})
\Bigg]
\Bigg\}
\,.
\label{eq:AveragedP2}
\end{align}

It is instructive to consider some simple cases allowing analytic treatment. First, we can recover the leading non-zero contribution in the opacity expansion by expanding (\ref{eq:AveragedP2}) in powers of the coupling $g^2$ (which enters in (\ref{eq:AveragedP2}) through $\mathcal{V}$)
\begin{align}
\left.\left\langle p_\bot^{2k} \vec{p} \right\rangle^{(1)}\right|_{N=1}&=
-\frac{\vec{u}}{2}\,\frac{f'(E)}{f(E)}\left.\left\langle p_\bot^{2k+2} \right\rangle\right|_{N=1}
-\frac{\vec{u}}{2E}L\int\frac{d^2\vec{p}\,d^2\vec{r}}{(2\pi)^2}\, p_\bot^{2k}\,e^{-i\vec{p}\cdot\vec{r}}\na_\b\na_\g\mathcal{V}_{\b\g}(\vec{r})\notag\\
&=
-\frac{\vec{u}}{2E}\,\mathcal{C}\r L\int\,
\frac{d^2\vec{p}}{(2\p)^2}\,p_\bot^{2k+2}\,
\left[E\frac{f'(E)}{f(E)}\,v^2(p_\bot)+
p_\bot^{2}\,\frac{\pa v^2}{\pa p^2_\bot}\right]\,,
\label{eq:op1p2k1}
\end{align}
which agrees with the results in \cite{Sadofyev:2021ohn}, as expected. Focusing on the particular value $k=0$, we notice that the result at first order in opacity is exact. Indeed, since $\mathcal{V}(\vec{r})$, $\na_\a\mathcal{V}(\vec{r})$, and $\mathcal{V}_{\a\b}(\vec{r})$ are zero at $\vec{r}=\vec{0}$, we find that for $k=0$ the all-order result in (\ref{eq:AveragedP2}) yields
\begin{align}
\left\langle \vec{p} \right\rangle^{(1)}&=-\frac{\vec{u}}{2}\frac{f'(E)}{f(E)}\left\langle p_\bot^{2} \right\rangle
+\frac{\vec{u}}{2E}\na_\b\Bigg\{e^{-\mathcal{V}(\vec{r})L}\Bigg[
L^2\Big(\mathcal{V}(\vec{r})\delta_{\b\g} 
+ \mathcal{V}_{\b\g}(\vec{r})\Big)\na_\g\mathcal{V}(\vec{r})
-L\na_\g\mathcal{V}_{\b\g}(\vec{r})
\Bigg]\Bigg\}\Bigg|_{\vec{r}=\vec{0}}\notag\\
&=-\frac{\vec{u}}{2E}\mathcal{C}\r L\int\,\frac{d^2\vec{p}}{(2\p)^2}\,p_\bot^{2}\,
\left[E\frac{f'(E)}{f(E)}\,v^2(p_\bot)+p_\bot^{2}\,\frac{\pa v^2}{\pa p^2_\bot}\right]
\end{align}
which agrees with (\ref{eq:op1p2k1}) for $k=0$. In general, for integer $k>0$, the opacity series up to $k+1$-th order is necessary to obtain the exact all-order result. 

The momentum integrals in \eqref{AveragedP} are well-defined for $k>-1$. However, we are also interested in comparing with the result for $k=-1$ obtained at leading order in opacity in \cite{Sadofyev:2021ohn}, and thus we have to be careful with their regularization. Since the additional insertions of $\mathcal{V}$ in \eqref{AveragedP} make the momentum integrals more convergent, the only possibly divergent term in this equation is the zeroth order in opacity (or equivalently, the $L=0$ contribution). This contribution is expected to disappear though, since in the absence of matter the broadening effects are controlled by the initial distribution (\ref{N0delta}) and one has to regulate (\ref{AveragedP}) accordingly. Indeed, if we take $L=0$ in (\ref{eq:momdist}), then up to the first subeikonal order the corresponding moment reads 
\begin{align}
\left\langle \frac{\tvec{p}}{p_\bot^2} \right\rangle\Bigg|_{L=0}=
\frac{1}{\N}\int\frac{d^2\vec{p}\,d^2\vec{p}_0}{(2\pi)^2}
\,\frac{\vec{p}}{p_\bot^2}
\,\d^{(2)}(\vec{p}-\vec{p}_0)
\,\left[1-\tvec{u}\cdot\left(\tvec{p}-\tvec{p}_0\right)\frac{\pa}{\pa E}\right]\, 
\frac{d\N^{(0)}}{d^2\tvec{p}_0dE}\,,
\label{eq:Lequal0}
\end{align}
where $\N$ is a normalization factor. Assuming that the initial distribution is isotropic, and has a small but finite width, we find that $\left\langle \tvec{p}/p_\bot^2 \right\rangle\Big|_{L=0}=0$ after angular integration. 

To first order in opacity, all momentum integrals are well defined and \eqref{eq:op1p2k1} is still valid. Using the explicit form of $v$ from the GW model given in \eqref{eq:potential}, we get
\begin{align}
\left.\left\langle \frac{\vec{p}}{p_\bot^2}\right\rangle^{(1)}\right|_{N=1} &= -\chi \,\frac{\vec{u}}{2}\left(\frac{f'(E)}{f(E)}-\frac{1}{E}\right)\, ,
\label{eq:firstopacity}
\end{align}
where we have introduced the opacity $\chi \equiv \mathcal{C}g^4\r L/(4\p \m^2)$. This result is, as expected, in full agreement with the outcome of \cite{Sadofyev:2021ohn}.

It is also interesting to compare $\left.\left\langle \vec{p}/p_\bot^2\right\rangle\right|_{N=1}$ with the full result accurate to all orders in opacity. In order to do so, we regularize $\left\langle p_\bot^{2k} \vec{p} \right\rangle$ for $k=-1$ in \eqref{AveragedP} with a finite mass
\begin{align}
\label{AveragedFiniteM}
\left\langle \frac{\vec{p}}{p_\bot^2+m^2}\right\rangle^{(1)}
&=-\frac{\vec{u}}{2}\,\frac{f'(E)}{f(E)}\int\frac{d^2\vec{p}\,d^2\vec{r}}{(2\pi)^2}\,
\frac{p^2_\bot}{p^2_\bot+m^2}\,
e^{-i\vec{p}\cdot\vec{r}}e^{-\mathcal{V}(\vec{r})L}
+\int\frac{d^2\vec{p}\,d^2\vec{r}}{(2\pi)^2}\, \frac{\vec{p}}{p^2_\bot+m^2}
\,e^{-i\vec{p}\cdot\vec{r}}
\notag\\
&\hspace{0cm}\times e^{-\mathcal{V}(\vec{r})L}\,\frac{u_\a}{E}\,
\Bigg[iL^2\,\Big(\mathcal{V}(\vec{r})\delta_{\a\b} 
+ \mathcal{V}_{\a\b}(\vec{r})\Big)\na_\b\mathcal{V}(\vec{r})
-iL\,\na_\b\mathcal{V}_{\a\b}(\vec{r})
\Bigg]
\,,
\end{align}
which can be numerically evaluated. Setting $f(E)\propto E^{-n}$ and removing dimensionful factors in \eqref{eq:AveragedP2}, we find
\begin{figure}[t!]
    \hspace{-4cm}\includegraphics[width=0.85\textwidth]{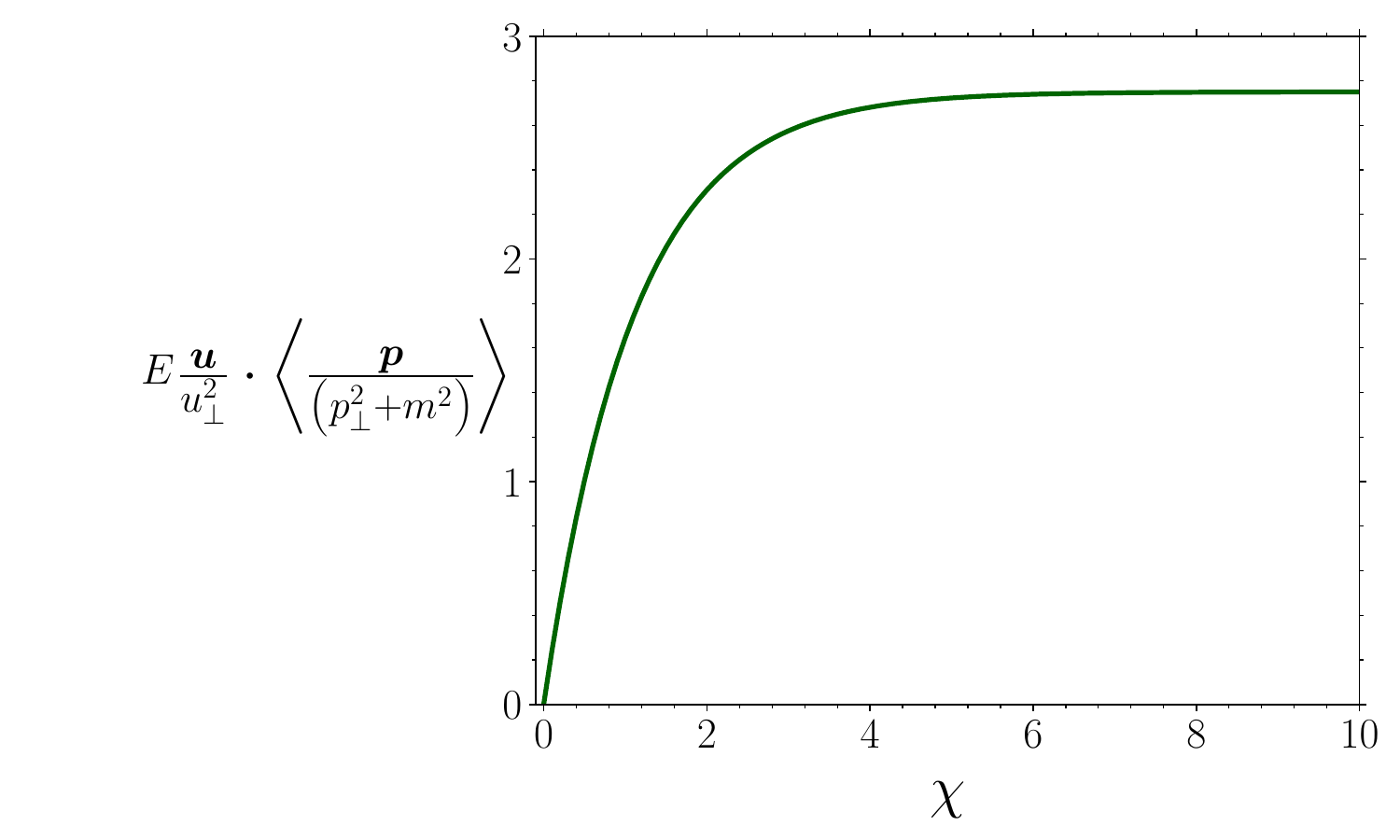}
    \caption{Numerical evaluation of (\ref{eq:Avkm1dimless}) for $n=4$ and $m=0.01 \m$ as a function of $\chi$. Note that the $L=0$ contribution given in (\ref{eq:Lequal0}) has been subtracted.}
\label{Fig2}
\end{figure}
\begin{align}
E\frac{\vec{u}}{u_\bot^2}\cdot\left\langle\frac{\vec{p}}{p_\bot^2+m^2}\right\rangle &= \frac{n}{2}\int\frac{d^2\vec{p}\,d^2\vec{r}}{(2\pi)^2}\,
\frac{p^2_\bot}{p^2_\bot+m^2}\,
e^{-i\vec{p}\cdot\vec{r}}e^{-\mathcal{V}(\vec{r})L}-\frac{i}{2}\int\frac{d^2\vec{p}\,d^2\vec{r}}{(2\pi)^2}\, \frac{1}{p_\perp^2+m^2}\,e^{-i\vec{p}\cdot\vec{r}}
\notag\\
&\hspace{-0.5cm}
\times \na_\a
\Bigg\{e^{-\mathcal{V}(\vec{r})L}
\Bigg[iL^2\,
\Big(\mathcal{V}(\vec{r})\delta_{\a\b} + \mathcal{V}_{\a\b}(\vec{r})\Big)\na_\b\mathcal{V}(\vec{r})
-iL\,\na_\b\mathcal{V}_{\a\b}(\vec{r})
\Bigg]
\Bigg\}
\,,
\label{eq:Avkm1dimless}
\end{align}
which is independent of $E$ and $\vec{u}$. Notice also that we have omitted the superscript for brevity.

We illustrate the full dependence of the all order result on the opacity variable $\chi$ in Fig.~\ref{Fig2} where we present the numerical evaluation of \eqref{eq:Avkm1dimless} for $n=4$, with $m$ sufficiently small, and with the $L=0$ contribution subtracted for stability. The curve clearly shows a linear behavior for small values of $\chi$, consistent with the $N=1$ result in Eq.~(\ref{eq:firstopacity}). For larger values of $\chi$, however, the curve saturates. This latter property is not present at any fixed order in the opacity expansion, thus showing how the resummed picture is essential to describe some observables in dense nuclear matter.

\section{Conclusions and outlook}
\label{sec:conclusions}

In this paper, we study how jets interact with homogeneous flowing QCD matter accounting for multiple interactions of the leading parton with the medium. By combining the leading subeikonal terms obtained at first order in opacity in \cite{Sadofyev:2021ohn} with the resummation techniques of the BDMPS-Z framework (which resums all orders in opacity in the eikonal limit), we succeed in finding the leading subeikonal corrections to the in-medium parton propagation to all orders in opacity. This is achieved at the amplitude level by deriving a single-particle in-medium propagator, see \eqref{G0} and \eqref{G1}. This result is then used to obtain the jet broadening distribution by computing the medium average of two propagators. Having a preferred direction set by the flow velocity clearly breaks azimuthal symmetry, and thus induces an anisotropy in the corresponding momentum distributions. In order to quantify and illustrate this effect, we have obtained the odd moments of the final state distribution $\langle p_\bot^{2k} \vec{p}\rangle$, which, contrary to the static case, do not vanish, and thus serve as the simplest probe of flow effects in jet-medium interactions. As expected, our all-order in opacity result agrees with the first order case examined in \cite{Sadofyev:2021ohn} in the small opacity limit. In particular we have analysed two specific cases: $k=0$ where the first order in opacity gives the exact (all-order) result, and $k=-1$ where a complete description can only be achieved when all orders in opacity are resummed.

Now that we have an expression for the in-medium propagator in the presence of flow and know how to compute medium averages of products of these propagators, the next natural step would be to derive the medium-induced radiation spectrum to all orders in opacity (including flow). This will be addressed in a separate publication. In principle, one would expect in-medium radiation, and more generally jet substructure observables, to be more sensitive to the transverse dynamics of the medium as daughter partons are less energetic. Further analyses on jet observables, in the direction of \cite{Antiporda:2021hpk} and extensions to the all-order in opacity calculations, will thus provide additional tools for studying the medium flow.

It is worth emphasizing that the formalism developed here can in principle be adapted to be used in general nuclear matter (hot or cold), where, in some cases, modeling the medium as a background field, like in Eq.~\eqref{eq:A_mu}, is still a reasonable approximation. Incorporating the effects of collective motion to jet observables for small collision systems may help elucidating the origin of the non-zero flow harmonics in proton-ion collisions or the recent measurements on high-$p_T$ jet-$v_2$. In this direction, it would be interesting to compare the results arising from our formalism both to recent developments in the subeikonal expansion in the CGC context \cite{Altinoluk:2020oyd,Altinoluk:2021lvu}, and to calculations of broadening in the glasma phase in the early stages of heavy ion collisions yielding to sizable anisotropies \cite{Ipp:2020mjc,Hauksson:2021okc,Carrington:2022bnv}. Similarly, one may attempt applying the developed description in the context of probe-matter interaction in deep inelastic scattering, see e.g. \cite{Wang:2001ifa,Zhang:2003wk, Sievert:2018imd,Arratia:2019vju,Li:2020zbk,Li:2020rqj,Arratia:2020nxw,Zhang:2021tcc,Sadofyev:2022hhw}, where the anisotropy is caused for instance by collective motion of nucleons. 

Finally, it would also be interesting to gain insight into the sensitivity of medium evolution effects to the strength of the interactions. In the case of strongly coupled holographic models, the effects due to flow, hydrodynamic gradients, and background fields have been studied for a wide variety of hard probes \cite{Lekaveckas:2013lha, Rajagopal:2015roa, Sadofyev:2015hxa, Li:2016bbh, Reiten:2019fta, Arefeva:2020jvo, Casalderrey-Solana:2014bpa,Rajagopal:2016uip, Brewer:2017fqy, Brewer:2018mpk}. The framework developed in this paper, along with the results in \cite{Sadofyev:2021ohn, Barata:2022krd}, could be used to provide the weak coupling counterpart of these analyses.

\acknowledgements
The authors would like to thank N. Armesto, J. Barata, X. Mayo, M. Sievert, and I. Vitev for discussions and comments on this work. This work is supported by European Research Council project ERC-2018-ADG-835105 YoctoLHC; by Maria de Maetzu excellence program under project MDM-2016-0692 and CEX2020-001035-M; by Spanish Research State Agency under project PID2020-119632GB-I00; and by Xunta de Galicia (Centro singular de investigación de Galicia accreditation 2019-2022), by European Union ERDF. A.V.S. has received funding from the European Union’s Horizon 2020 research and innovation program under the Marie Sklodowska-Curie grant agreement No 101032858 (JetT). C.A. has received funding from the European Union’s Horizon 2020 research and innovation program under the Marie Sklodowska-Curie grant agreement No 893021 (JQ4LHC).

\appendix

\section{The harmonic oscillator approximation}

\label{ap:A}

The harmonic approximation is easily formulated in coordinate space, where the interaction potential is:
\be
{\cal V}(\vec{r}) = \frac{1}{4}\hat q r_\bot^2\,.
\ee
We thus have
\be
{\cal V}_{\alpha\beta}(\vec{r}) = -\frac{1}{4}\hat q\left(\frac{\delta_{\alpha\beta}}{2} r_\bot^2 + \vec{r}_\alpha\vec{r}_\beta\right)\,.
\ee

We can now go back to the expressions for the broadening probability in section~\ref{sec:broadening} and evaluate them for this particular approximation. Eq.~\eqref{eq:Pzero} then takes the form
\be
{\cal P}^{(0)}(\vec{r},L;\vec{r}_0,0) = e^{-\frac{1}{4}\hat q r_\bot^2L}\delta^{(2)}(\vec{r}-\vec{r}_0),
\ee
with Fourier transform
\be
{\cal P}^{(0)}(\vec{p},L;\vec{p}_0,0) = \frac{4\pi}{\hat qL}e^{-\frac{(\vec{p}-\vec{p}_0)^2}{\hat qL}}\,,
\ee
and Eq.~\eqref{eq:P1coor}
\begin{align}
\mathcal{P}^{(1)}(\vec{r},L;\vec{r}_0,0) &= e^{-\frac{1}{4}\hat qr_\bot^2L}\,\frac{u_\a}{E}\left\{-\frac{i}{2}\hat qL\,\left(\frac{\delta_{\alpha\beta}}{2}r_\bot^2-\vec{r}_\alpha\vec{r}_\beta\right)\na_\b\,\delta^{(2)}(\vec{r}-\vec{r}_0)\right.
\nonumber\\
&\quad\left.+\,i\hat qL\vec{r}_\alpha\left(1-\frac{1}{16}\hat qLr_\bot^2\right)\delta^{(2)}(\vec{r}-\vec{r}_0)\right\}\,.
\end{align}
Fourier transforming to momentum space, we get from Eq.~\eqref{eq:P1withp0}
\begin{align}
\mathcal{P}^{(1)}(\vec{p},L;\vec{p}_0,0) &= 
\int d^2\vec{r}\, e^{-i(\vec{p}-\vec{p}_0)\cdot\vec{r}}e^{-\frac{1}{4}\hat q r_\bot^2 L}\,\frac{u_\a}{E}\,
\left[\frac{1}{2}\hat qL\,\vec{p}_{0\b}\,\left(\frac{\delta_{\alpha\beta}}{2}r_\bot^2-\vec{r}_\alpha\vec{r}_\beta\right)\right. \nonumber\\
&\left.\quad+\,i\hat q L\,\vec{r}_\alpha\left(1-\frac{1}{16}\hat qLr_\bot^2\right)\right]\,.
\end{align}
All the additional factors of $\vec{r}$ appearing inside the square brackets can be replaced by derivatives with respect to $\vec{p}$, yielding
\begin{align}
\mathcal{P}^{(1)}(\vec{p},L;\vec{p}_0,0) &= 2\pi\, \frac{u_\a}{E}\,\left[-\vec{p}_{0\b}\,\left(\frac{\delta_{\alpha\beta}}{2}\nabla_p^2-\nabla_p^\alpha\nabla_p^\beta\right) - \left(2+\frac{1}{8}\hat qL\nabla_p^2\right)\nabla_p^\alpha\right]e^{-\frac{(\vec{p}-\vec{p}_0)^2}{\hat qL}}\nonumber\\
&  \hspace{-2cm} = \frac{2\p}{\hat{q}^2 L^2} \frac{u_\a}{E} e^{-\frac{(\vec{p}-\vec{p}_0)^2}{\hat{q} L}}\left[(p-p_0)_\a\left(p_\perp^2+2\vec{p}\cdot\vec{p}_0-3p_{0\perp}^2+2\hat{q} L\right)-2p_{0\a}(\vec{p}-\vec{p}_0)^2\right]
\end{align}

\bibliographystyle{bibstyle}
\bibliography{references}

\begin{thebibliography}{10}
\ifx\href\asklfhas\newcommand{\href}[2]{#2}\fi
\ifx\arxivref\asklfhas\newcommand{\arxivref}[2]{\href{http://arxiv.org/abs/#1}{#2}}\fi
\ifx\doiref\asklfhas\newcommand{\doiref}[2]{\href{http://dx.doi.org/#1}{#2}}\fi
\parskip 0pt
\normalsize

\bibitem{PHENIX:2001hpc}
PHENIX Collaboration, K.~Adcox et~al.,
\textit{``{Suppression of hadrons with large transverse momentum in central
  Au+Au collisions at $\sqrt{s_{NN}}$ = 130-GeV}''},
\doiref{10.1103/PhysRevLett.88.022301}{Phys.~Rev.~Lett. \textbf{88}, 022301
  (2002)},
\normalsize{\texttt{\arxivref{nucl-ex/0109003}{nucl-ex/0109003}}}.

\bibitem{STAR:2002ggv}
STAR Collaboration, C.~Adler et~al.,
\textit{``{Centrality dependence of high $p_{T}$ hadron suppression in Au+Au
  collisions at $\sqrt{s}_{NN}$ = 130-GeV}''},
\doiref{10.1103/PhysRevLett.89.202301}{Phys.~Rev.~Lett. \textbf{89}, 202301
  (2002)},
\normalsize{\texttt{\arxivref{nucl-ex/0206011}{nucl-ex/0206011}}}.

\bibitem{Vitev:2002pf}
I.~Vitev \& M.~Gyulassy,
\textit{``{High $p_{T}$ tomography of $d$ + Au and Au+Au at SPS, RHIC, and
  LHC}''},
\doiref{10.1103/PhysRevLett.89.252301}{Phys.~Rev.~Lett. \textbf{89}, 252301
  (2002)},
\normalsize{\texttt{\arxivref{hep-ph/0209161}{hep-ph/0209161}}}.

\bibitem{Majumder:2006wi}
A.~Majumder, B.~Muller \& S.~A. Bass,
\textit{``{Longitudinal Broadening of Quenched Jets in Turbulent Color
  Fields}''},
\doiref{10.1103/PhysRevLett.99.042301}{Phys.~Rev.~Lett. \textbf{99}, 042301
  (2007)},
\normalsize{\texttt{\arxivref{hep-ph/0611135}{hep-ph/0611135}}}.

\bibitem{Xu:2014ica}
J.~Xu, A.~Buzzatti \& M.~Gyulassy,
\textit{``{Azimuthal jet flavor tomography with CUJET2.0 of nuclear collisions
  at RHIC and LHC}''},
\doiref{10.1007/JHEP08(2014)063}{JHEP \textbf{1408}, 063 (2014)},
\normalsize{\texttt{\arxivref{1402.2956}{arXiv:1402.2956}}}.

\bibitem{Djordjevic:2016vfo}
M.~Djordjevic, B.~Blagojevic \& L.~Zivkovic,
\textit{``{Mass tomography at different momentum ranges in quark-gluon
  plasma}''},
\doiref{10.1103/PhysRevC.94.044908}{Phys.~Rev.~C \textbf{94}, 044908 (2016)},
\normalsize{\texttt{\arxivref{1601.07852}{arXiv:1601.07852}}}.

\bibitem{Apolinario:2017sob}
L.~Apolin\'ario, J.~G. Milhano, G.~P. Salam \& C.~A. Salgado,
\textit{``{Probing the time structure of the quark-gluon plasma with top
  quarks}''},
\doiref{10.1103/PhysRevLett.120.232301}{Phys.~Rev.~Lett. \textbf{120}, 232301
  (2018)},
\normalsize{\texttt{\arxivref{1711.03105}{arXiv:1711.03105}}}.

\bibitem{Andres:2019eus}
C.~Andres, N.~Armesto, H.~Niemi, R.~Paatelainen \& C.~A. Salgado,
\textit{``{Jet quenching as a probe of the initial stages in heavy-ion
  collisions}''},
\doiref{10.1016/j.physletb.2020.135318}{Phys.~Lett.~B \textbf{803}, 135318
  (2020)},
\normalsize{\texttt{\arxivref{1902.03231}{arXiv:1902.03231}}}.

\bibitem{Feal:2019xfl}
X.~Feal, C.~A. Salgado \& R.~A. Vazquez,
\textit{``{Jet quenching test of the QCD matter created at RHIC and the LHC
  needs opacity-resummed medium induced radiation}''},
\doiref{10.1016/j.physletb.2021.136251}{Phys.~Lett.~B \textbf{816}, 136251
  (2021)},
\normalsize{\texttt{\arxivref{1911.01309}{arXiv:1911.01309}}}.

\bibitem{He:2020iow}
Y.~He, L.-G. Pang \& X.-N. Wang,
\textit{``{Gradient Tomography of Jet Quenching in Heavy-Ion Collisions}''},
\doiref{10.1103/PhysRevLett.125.122301}{Phys.~Rev.~Lett. \textbf{125}, 122301
  (2020)},
\normalsize{\texttt{\arxivref{2001.08273}{arXiv:2001.08273}}}.

\bibitem{Apolinario:2020uvt}
L.~Apolin\'ario, A.~Cordeiro \& K.~Zapp,
\textit{``{Time reclustering for jet quenching studies}''},
\doiref{10.1140/epjc/s10052-021-09346-8}{Eur.~Phys.~J.~C \textbf{81}, 561
  (2021)},
\normalsize{\texttt{\arxivref{2012.02199}{arXiv:2012.02199}}}.

\bibitem{Sadofyev:2021ohn}
A.~V. Sadofyev, M.~D. Sievert \& I.~Vitev,
\textit{``{Ab~initio coupling of jets to collective flow in the opacity
  expansion approach}''},
\doiref{10.1103/PhysRevD.104.094044}{Phys.~Rev.~D \textbf{104}, 094044 (2021)},
\normalsize{\texttt{\arxivref{2104.09513}{arXiv:2104.09513}}}.

\bibitem{Du:2021pqa}
Y.-L. Du, D.~Pablos \& K.~Tywoniuk,
\textit{``{Jet tomography in heavy ion collisions with deep learning}''},
\normalsize{\texttt{\arxivref{2106.11271}{arXiv:2106.11271}}}.

\bibitem{Antiporda:2021hpk}
L.~Antiporda, J.~Bahder, H.~Rahman \& M.~D. Sievert,
\textit{``{Jet Drift and Collective Flow in Heavy-Ion Collisions}''},
\normalsize{\texttt{\arxivref{2110.03590}{arXiv:2110.03590}}}.

\bibitem{Hauksson:2021okc}
S.~Hauksson, S.~Jeon \& C.~Gale,
\textit{``{The momentum broadening of energetic partons in an anisotropic
  plasma}''},
\normalsize{\texttt{\arxivref{2109.04575}{arXiv:2109.04575}}}.

\bibitem{Barata:2022krd}
J.~Barata, A.~V. Sadofyev \& C.~A. Salgado,
\textit{``{Jet broadening in dense inhomogeneous matter}''},
\normalsize{\texttt{\arxivref{2202.08847}{arXiv:2202.08847}}}.

\bibitem{Fu:2022idl}
Y.~Fu, J.~Casalderrey-Solana \& X.-N. Wang,
\textit{``{Asymmetric transverse momentum broadening in an inhomogeneous
  medium}''},
\normalsize{\texttt{\arxivref{2204.05323}{arXiv:2204.05323}}}.

\bibitem{Sadofyev:2022hhw}
A.~V. Sadofyev, M.~D. Sievert \& I.~Vitev,
\textit{``{Jets in evolving matter within the opacity expansion approach}''},
\doiref{10.21468/SciPostPhysProc.8.046}{SciPost~Phys.~Proc. \textbf{8}, 046
  (2022)},
\normalsize{\texttt{\arxivref{2207.07679}{arXiv:2207.07679}}}.

\bibitem{Baier:1996kr}
R.~Baier, Y.~L. Dokshitzer, A.~H. Mueller, S.~Peigne \& D.~Schiff,
\textit{``{Radiative energy loss of high-energy quarks and gluons in a finite
  volume quark - gluon plasma}''},
\doiref{10.1016/S0550-3213(96)00553-6}{Nucl.~Phys.~B \textbf{483}, 291 (1997)},
\normalsize{\texttt{\arxivref{hep-ph/9607355}{hep-ph/9607355}}}.

\bibitem{Baier:1996sk}
R.~Baier, Y.~L. Dokshitzer, A.~H. Mueller, S.~Peigne \& D.~Schiff,
\textit{``{Radiative energy loss and p(T) broadening of high-energy partons in
  nuclei}''},
\doiref{10.1016/S0550-3213(96)00581-0}{Nucl.~Phys. \textbf{B484}, 265 (1997)},
\normalsize{\texttt{\arxivref{hep-ph/9608322}{hep-ph/9608322}}}.

\bibitem{Zakharov:1996fv}
B.~G. Zakharov,
\textit{``{Fully quantum treatment of the Landau-Pomeranchuk-Migdal effect in
  QED and QCD}''},
\doiref{10.1134/1.567126}{JETP~Lett. \textbf{63}, 952 (1996)},
\normalsize{\texttt{\arxivref{hep-ph/9607440}{hep-ph/9607440}}}.

\bibitem{Zakharov:1997uu}
B.~G. Zakharov,
\textit{``{Radiative energy loss of high-energy quarks in finite size nuclear
  matter and quark - gluon plasma}''},
\doiref{10.1134/1.567389}{JETP~Lett. \textbf{65}, 615 (1997)},
\normalsize{\texttt{\arxivref{hep-ph/9704255}{hep-ph/9704255}}}.

\bibitem{Gyulassy:2000fs}
M.~Gyulassy, P.~Levai \& I.~Vitev,
\textit{``{NonAbelian energy loss at finite opacity}''},
\doiref{10.1103/PhysRevLett.85.5535}{Phys.~Rev.~Lett. \textbf{85}, 5535
  (2000)},
\normalsize{\texttt{\arxivref{nucl-th/0005032}{nucl-th/0005032}}}.

\bibitem{Gyulassy:2000er}
M.~Gyulassy, P.~Levai \& I.~Vitev,
\textit{``{Reaction operator approach to nonAbelian energy loss}''},
\doiref{10.1016/S0550-3213(00)00652-0}{Nucl.~Phys. \textbf{B594}, 371 (2001)},
\normalsize{\texttt{\arxivref{nucl-th/0006010}{nucl-th/0006010}}}.

\bibitem{Wiedemann:2000za}
U.~A. Wiedemann,
\textit{``{Gluon radiation off hard quarks in a nuclear environment: Opacity
  expansion}''},
\doiref{10.1016/S0550-3213(00)00457-0}{Nucl.~Phys.~B \textbf{588}, 303 (2000)},
\normalsize{\texttt{\arxivref{hep-ph/0005129}{hep-ph/0005129}}}.

\bibitem{Wang:2001ifa}
X.-N. Wang \& X.-f. Guo,
\textit{``{Multiple parton scattering in nuclei: Parton energy loss}''},
\doiref{10.1016/S0375-9474(01)01130-7}{Nucl.~Phys. \textbf{A696}, 788 (2001)},
\normalsize{\texttt{\arxivref{hep-ph/0102230}{hep-ph/0102230}}}.

\bibitem{Arnold:2002ja}
P.~B. Arnold, G.~D. Moore \& L.~G. Yaffe,
\textit{``{Photon and gluon emission in relativistic plasmas}''},
\doiref{10.1088/1126-6708/2002/06/030}{JHEP \textbf{0206}, 030 (2002)},
\normalsize{\texttt{\arxivref{hep-ph/0204343}{hep-ph/0204343}}}.

\bibitem{Djordjevic:2003zk}
M.~Djordjevic \& M.~Gyulassy,
\textit{``{Heavy quark radiative energy loss in QCD matter}''},
\doiref{10.1016/j.nuclphysa.2003.12.020}{Nucl.~Phys. \textbf{A733}, 265
  (2004)},
\normalsize{\texttt{\arxivref{nucl-th/0310076}{nucl-th/0310076}}}.

\bibitem{Mehtar-Tani:2006vpj}
Y.~Mehtar-Tani,
\textit{``{Relating the description of gluon production in pA collisions and
  parton energy loss in AA collisions}''},
\doiref{10.1103/PhysRevC.75.034908}{Phys.~Rev.~C \textbf{75}, 034908 (2007)},
\normalsize{\texttt{\arxivref{hep-ph/0606236}{hep-ph/0606236}}}.

\bibitem{Caron-Huot:2010qjx}
S.~Caron-Huot \& C.~Gale,
\textit{``{Finite-size effects on the radiative energy loss of a fast parton in
  hot and dense strongly interacting matter}''},
\doiref{10.1103/PhysRevC.82.064902}{Phys.~Rev.~C \textbf{82}, 064902 (2010)},
\normalsize{\texttt{\arxivref{1006.2379}{arXiv:1006.2379}}}.

\bibitem{Sievert:2018imd}
M.~D. Sievert \& I.~Vitev,
\textit{``{Quark branching in QCD matter to any order in opacity beyond the
  soft gluon emission limit}''},
\doiref{10.1103/PhysRevD.98.094010}{Phys.~Rev. \textbf{D98}, 094010 (2018)},
\normalsize{\texttt{\arxivref{1807.03799}{arXiv:1807.03799}}}.

\bibitem{Andres:2020vxs}
C.~Andres, L.~Apolin\'ario \& F.~Dominguez,
\textit{``{Medium-induced gluon radiation with full resummation of multiple
  scatterings for realistic parton-medium interactions}''},
\doiref{10.1007/JHEP07(2020)114}{JHEP \textbf{2007}, 114 (2020)},
\normalsize{\texttt{\arxivref{2002.01517}{arXiv:2002.01517}}}.

\bibitem{Barata:2021wuf}
J.~Barata, Y.~Mehtar-Tani, A.~Soto-Ontoso \& K.~Tywoniuk,
\textit{``{Medium-induced radiative kernel with the Improved Opacity
  Expansion}''},
\normalsize{\texttt{\arxivref{2106.07402}{arXiv:2106.07402}}}.

\bibitem{Casalderrey-Solana:2007knd}
J.~Casalderrey-Solana \& C.~A. Salgado,
\textit{``{Introductory lectures on jet quenching in heavy ion collisions}''},
Acta~Phys.~Polon.~B \textbf{38}, 3731 (2007),
\normalsize{\texttt{\arxivref{0712.3443}{arXiv:0712.3443}}}.

\bibitem{Blaizot:2012fh}
J.-P. Blaizot, F.~Dominguez, E.~Iancu \& Y.~Mehtar-Tani,
\textit{``{Medium-induced gluon branching}''},
\doiref{10.1007/JHEP01(2013)143}{JHEP \textbf{1301}, 143 (2013)},
\normalsize{\texttt{\arxivref{1209.4585}{arXiv:1209.4585}}}.

\bibitem{Wiedemann:2000ez}
U.~A. Wiedemann,
\textit{``{Transverse dynamics of hard partons in nuclear media and the QCD
  dipole}''},
\doiref{10.1016/S0550-3213(00)00286-8}{Nucl.~Phys.~B \textbf{582}, 409 (2000)},
\normalsize{\texttt{\arxivref{hep-ph/0003021}{hep-ph/0003021}}}.

\bibitem{Baier:1998yf}
R.~Baier, Y.~L. Dokshitzer, A.~H. Mueller \& D.~Schiff,
\textit{``{Radiative energy loss of high-energy partons traversing an expanding
  QCD plasma}''},
\doiref{10.1103/PhysRevC.58.1706}{Phys.~Rev.~C \textbf{58}, 1706 (1998)},
\normalsize{\texttt{\arxivref{hep-ph/9803473}{hep-ph/9803473}}}.

\bibitem{Gyulassy:2000gk}
M.~Gyulassy, I.~Vitev \& X.~Wang,
\textit{``{High p(T) azimuthal asymmetry in noncentral A+A at RHIC}''},
\doiref{10.1103/PhysRevLett.86.2537}{Phys.~Rev.~Lett. \textbf{86}, 2537
  (2001)},
\normalsize{\texttt{\arxivref{nucl-th/0012092}{nucl-th/0012092}}}.

\bibitem{Gyulassy:2001kr}
M.~Gyulassy, I.~Vitev, X.-N. Wang \& P.~Huovinen,
\textit{``{Transverse expansion and high p(T) azimuthal asymmetry at RHIC}''},
\doiref{10.1016/S0370-2693(02)01157-7}{Phys.~Lett.~B \textbf{526}, 301 (2002)},
\normalsize{\texttt{\arxivref{nucl-th/0109063}{nucl-th/0109063}}}.

\bibitem{Armesto:2004pt}
N.~Armesto, C.~A. Salgado \& U.~A. Wiedemann,
\textit{``{Measuring the collective flow with jets}''},
\doiref{10.1103/PhysRevLett.93.242301}{Phys.~Rev.~Lett. \textbf{93}, 242301
  (2004)},
\normalsize{\texttt{\arxivref{hep-ph/0405301}{hep-ph/0405301}}}.

\bibitem{Armesto:2004vz}
N.~Armesto, C.~A. Salgado \& U.~A. Wiedemann,
\textit{``{Low-p(T) collective flow induces high-p(T) jet quenching}''},
\doiref{10.1103/PhysRevC.72.064910}{Phys.~Rev.~C \textbf{72}, 064910 (2005)},
\normalsize{\texttt{\arxivref{hep-ph/0411341}{hep-ph/0411341}}}.

\bibitem{Baier:2006pt}
R.~Baier, A.~H. Mueller \& D.~Schiff,
\textit{``{How does transverse (hydrodynamic) flow affect jet-broadening and
  jet-quenching ?}''},
\doiref{10.1016/j.physletb.2007.03.048}{Phys.~Lett. \textbf{B649}, 147 (2007)},
\normalsize{\texttt{\arxivref{nucl-th/0612068}{nucl-th/0612068}}}.

\bibitem{Liu:2006he}
H.~Liu, K.~Rajagopal \& U.~A. Wiedemann,
\textit{``{Wilson loops in heavy ion collisions and their calculation in
  AdS/CFT}''},
\doiref{10.1088/1126-6708/2007/03/066}{JHEP \textbf{0703}, 066 (2007)},
\normalsize{\texttt{\arxivref{hep-ph/0612168}{hep-ph/0612168}}}.

\bibitem{Renk:2006sx}
T.~Renk, J.~Ruppert, C.~Nonaka \& S.~A. Bass,
\textit{``{Jet-quenching in a 3D hydrodynamic medium}''},
\doiref{10.1103/PhysRevC.75.031902}{Phys.~Rev. \textbf{C75}, 031902 (2007)},
\normalsize{\texttt{\arxivref{nucl-th/0611027}{nucl-th/0611027}}}.

\bibitem{Gyulassy:1993hr}
M.~Gyulassy \& X.-n. Wang,
\textit{``{Multiple collisions and induced gluon Bremsstrahlung in QCD}''},
\doiref{10.1016/0550-3213(94)90079-5}{Nucl.~Phys. \textbf{B420}, 583 (1994)},
\normalsize{\texttt{\arxivref{nucl-th/9306003}{nucl-th/9306003}}}.

\bibitem{Blaizot:2015lma}
J.-P. Blaizot \& Y.~Mehtar-Tani,
\textit{``{Jet Structure in Heavy Ion Collisions}''},
\doiref{10.1142/S021830131530012X}{Int.~J.~Mod.~Phys.~E \textbf{24}, 1530012
  (2015)},
\normalsize{\texttt{\arxivref{1503.05958}{arXiv:1503.05958}}}.

\bibitem{Isaksen:2020npj}
J.~H. Isaksen \& K.~Tywoniuk,
\textit{``{Wilson line correlators beyond the large-N$_{c}$}''},
\doiref{10.1007/JHEP11(2021)125}{JHEP \textbf{2021}, 125 (2020)},
\normalsize{\texttt{\arxivref{2107.02542}{arXiv:2107.02542}}}.

\bibitem{Altinoluk:2020oyd}
T.~Altinoluk, G.~Beuf, A.~Czajka \& A.~Tymowska,
\textit{``{Quarks at next-to-eikonal accuracy in the CGC: Forward quark-nucleus
  scattering}''},
\doiref{10.1103/PhysRevD.104.014019}{Phys.~Rev.~D \textbf{104}, 014019 (2021)},
\normalsize{\texttt{\arxivref{2012.03886}{arXiv:2012.03886}}}.

\bibitem{Altinoluk:2021lvu}
T.~Altinoluk \& G.~Beuf,
\textit{``{Quark and scalar propagators at next-to-eikonal accuracy in the CGC
  through a dynamical background gluon field}''},
\normalsize{\texttt{\arxivref{2109.01620}{arXiv:2109.01620}}}.

\bibitem{Ipp:2020mjc}
A.~Ipp, D.~I. M\"uller \& D.~Schuh,
\textit{``{Anisotropic momentum broadening in the 2+1D Glasma: analytic weak
  field approximation and lattice simulations}''},
\doiref{10.1103/PhysRevD.102.074001}{Phys.~Rev.~D \textbf{102}, 074001 (2020)},
\normalsize{\texttt{\arxivref{2001.10001}{arXiv:2001.10001}}}.

\bibitem{Carrington:2022bnv}
M.~E. Carrington, A.~Czajka \& S.~Mrowczynski,
\textit{``{Transport of hard probes through glasma}''},
\normalsize{\texttt{\arxivref{2202.00357}{arXiv:2202.00357}}}.

\bibitem{Zhang:2003wk}
B.-W. Zhang, E.~Wang \& X.-N. Wang,
\textit{``{Heavy quark energy loss in nuclear medium}''},
\doiref{10.1103/PhysRevLett.93.072301}{Phys.~Rev.~Lett. \textbf{93}, 072301
  (2004)},
\normalsize{\texttt{\arxivref{nucl-th/0309040}{nucl-th/0309040}}}.

\bibitem{Arratia:2019vju}
M.~Arratia, Y.~Song, F.~Ringer \& B.~V. Jacak,
\textit{``{Jets as precision probes in electron-nucleus collisions at the
  future Electron-Ion Collider}''},
\doiref{10.1103/PhysRevC.101.065204}{Phys.~Rev.~C \textbf{101}, 065204 (2020)},
\normalsize{\texttt{\arxivref{1912.05931}{arXiv:1912.05931}}}.

\bibitem{Li:2020zbk}
H.~T. Li, Z.~L. Liu \& I.~Vitev,
\textit{``{Heavy meson tomography of cold nuclear matter at the electron-ion
  collider}''},
\doiref{10.1016/j.physletb.2021.136261}{Phys.~Lett.~B \textbf{816}, 136261
  (2021)},
\normalsize{\texttt{\arxivref{2007.10994}{arXiv:2007.10994}}}.

\bibitem{Li:2020rqj}
H.~T. Li \& I.~Vitev,
\textit{``{Nuclear Matter Effects on Jet Production at Electron-Ion
  Colliders}''},
\doiref{10.1103/PhysRevLett.126.252001}{Phys.~Rev.~Lett. \textbf{126}, 252001
  (2021)},
\normalsize{\texttt{\arxivref{2010.05912}{arXiv:2010.05912}}}.

\bibitem{Arratia:2020nxw}
M.~Arratia, Z.-B. Kang, A.~Prokudin \& F.~Ringer,
\textit{``{Jet-based measurements of Sivers and Collins asymmetries at the
  future electron-ion collider}''},
\doiref{10.1103/PhysRevD.102.074015}{Phys.~Rev.~D \textbf{102}, 074015 (2020)},
\normalsize{\texttt{\arxivref{2007.07281}{arXiv:2007.07281}}}.

\bibitem{Zhang:2021tcc}
Y.-Y. Zhang \& X.-N. Wang,
\textit{``{Multiple parton scattering and gluon saturation in dijet production
  at EIC}''},
\normalsize{\texttt{\arxivref{2104.04520}{arXiv:2104.04520}}}.

\bibitem{Lekaveckas:2013lha}
M.~Lekaveckas \& K.~Rajagopal,
\textit{``{Effects of Fluid Velocity Gradients on Heavy Quark Energy Loss}''},
\doiref{10.1007/JHEP02(2014)068}{JHEP \textbf{1402}, 068 (2014)},
\normalsize{\texttt{\arxivref{1311.5577}{arXiv:1311.5577}}}.

\bibitem{Rajagopal:2015roa}
K.~Rajagopal \& A.~V. Sadofyev,
\textit{``{Chiral drag force}''},
\doiref{10.1007/JHEP10(2015)018}{JHEP \textbf{1510}, 018 (2015)},
\normalsize{\texttt{\arxivref{1505.07379}{arXiv:1505.07379}}}.

\bibitem{Sadofyev:2015hxa}
A.~V. Sadofyev \& Y.~Yin,
\textit{``{The charmonium dissociation in an “anomalous wind”}''},
\doiref{10.1007/JHEP01(2016)052}{JHEP \textbf{1601}, 052 (2016)},
\normalsize{\texttt{\arxivref{1510.06760}{arXiv:1510.06760}}}.

\bibitem{Li:2016bbh}
S.~Li, K.~A. Mamo \& H.-U. Yee,
\textit{``{Jet quenching parameter of the quark-gluon plasma in a strong
  magnetic field: Perturbative QCD and AdS/CFT correspondence}''},
\doiref{10.1103/PhysRevD.94.085016}{Phys.~Rev. \textbf{D94}, 085016 (2016)},
\normalsize{\texttt{\arxivref{1605.00188}{arXiv:1605.00188}}}.

\bibitem{Reiten:2019fta}
J.~Reiten \& A.~V. Sadofyev,
\textit{``{Drag force to all orders in gradients}''},
\doiref{10.1007/JHEP07(2020)146}{JHEP \textbf{2007}, 146 (2020)},
\normalsize{\texttt{\arxivref{1912.08816}{arXiv:1912.08816}}}.

\bibitem{Arefeva:2020jvo}
I.~Y. Aref'eva, A.~A. Golubtsova \& E.~Gourgoulhon,
\textit{``{Holographic drag force in 5d Kerr-AdS black hole}''},
\normalsize{\texttt{\arxivref{2004.12984}{arXiv:2004.12984}}}.

\bibitem{Casalderrey-Solana:2014bpa}
J.~Casalderrey-Solana, D.~C. Gulhan, J.~G. Milhano, D.~Pablos \& K.~Rajagopal,
\textit{``{A Hybrid Strong/Weak Coupling Approach to Jet Quenching}''},
\doiref{10.1007/JHEP09(2015)175}{JHEP \textbf{1410}, 019 (2014)},
\normalsize{\texttt{\arxivref{1405.3864}{arXiv:1405.3864}}},
[Erratum: JHEP 09, 175 (2015)].

\bibitem{Rajagopal:2016uip}
K.~Rajagopal, A.~V. Sadofyev \& W.~van~der~Schee,
\textit{``{Evolution of the jet opening angle distribution in holographic
  plasma}''},
\doiref{10.1103/PhysRevLett.116.211603}{Phys.~Rev.~Lett. \textbf{116}, 211603
  (2016)},
\normalsize{\texttt{\arxivref{1602.04187}{arXiv:1602.04187}}}.

\bibitem{Brewer:2017fqy}
J.~Brewer, K.~Rajagopal, A.~Sadofyev \& W.~Van Der~Schee,
\textit{``{Evolution of the Mean Jet Shape and Dijet Asymmetry Distribution of
  an Ensemble of Holographic Jets in Strongly Coupled Plasma}''},
\doiref{10.1007/JHEP02(2018)015}{JHEP \textbf{1802}, 015 (2018)},
\normalsize{\texttt{\arxivref{1710.03237}{arXiv:1710.03237}}}.

\bibitem{Brewer:2018mpk}
J.~Brewer, A.~Sadofyev \& W.~van~der~Schee,
\textit{``{Jet shape modifications in holographic dijet systems}''},
\doiref{10.1016/j.physletb.2021.136492}{Phys.~Lett.~B \textbf{820}, 136492
  (2021)},
\normalsize{\texttt{\arxivref{1809.10695}{arXiv:1809.10695}}}.

\end{thebibliography}

\end{document}